\documentclass[twolcolumn, superscriptaddress, prl, reprint]{revtex4-2}
\usepackage[latin9]{inputenc}
\usepackage{verbatim}
\usepackage{amsmath}
\usepackage{notes2bib}
\usepackage{amssymb}
\usepackage{graphicx}

\usepackage{amsthm}
\usepackage{fancyhdr}
\usepackage{epsfig}
\usepackage{bbm}
\usepackage{subfigure}
\usepackage{float}
\usepackage{xcolor}
\usepackage[normalem]{ulem}

\begin{document}
\title{Oscillatory force autocorrelations in equilibrium odd-diffusive systems}

\author{Erik Kalz}
\affiliation{University of Potsdam, Institute of Physics and Astronomy, D-14476 Potsdam, Germany}

\author{Hidde Derk Vuijk}
\affiliation{University of Augsburg, Institute of Physics, D-86159 Augsburg, Germany}

\author{Jens-Uwe Sommer}
\affiliation{Leibniz-Institute for Polymer Research, Institute Theory of Polymers, D-01069 Dresden, Germany}
\affiliation{Technical University of Dresden, Institute for Theoretical Physics, D-01069 Dresden, Germany}
\affiliation{Technical University of Dresden, Cluster of Excellence Physics of Life, D-01069 Dresden, Germany}

\author{Ralf Metzler}
\affiliation{University of Potsdam, Institute of Physics and Astronomy, D-14476 Potsdam, Germany}
\affiliation{Asia Pacific Centre for Theoretical Physics, KR-37673 Pohang, Republic of Korea}

\author{Abhinav Sharma}
\affiliation{University of Augsburg, Institute of Physics, D-86159 Augsburg, Germany}
\affiliation{Leibniz-Institute for Polymer Research, Institute Theory of Polymers, D-01069 Dresden, Germany}

\begin{abstract}
The force autocorrelation function (FACF), a concept of fundamental interest in statistical mechanics, encodes the effect of interactions on the dynamics of a tagged particle.
In equilibrium, the FACF is believed to decay monotonically in time which is a signature of slowing down of the dynamics of the tagged particle due to interactions. 
Here we analytically show that in odd-diffusive systems, which are characterized by a diffusion tensor with antisymmetric elements, the FACF can become negative and even exhibit temporal oscillations.
We also demonstrate that, despite the isotropy, the knowledge of FACF alone is not sufficient to describe the dynamics: the full autocorrelation tensor is required and contains an antisymmetric part.
These unusual properties translate into enhanced dynamics of the tagged particle quantified via the self-diffusion coefficient that, remarkably, increases due to particle interactions.

\end{abstract}
\maketitle


\textit{Introduction.} Time integrals of appropriate correlation functions are related to transport coefficients via the Green-Kubo relations~\cite{green1954markoff,kubo2012statistical}. The self-diffusion coefficient, for instance, is determined by the integral of the force autocorrelation function (FACF), which quantifies the effect of interactions on the diffusive dynamics of a particle. Previous studies have shown that in overdamped equilibrium systems, the FACF decays monotonically for all densities independently of the nature of interaction between particles~\cite{ hanna1981velocity,hanna1982self,sharma2016communication,mandal2019persistent}. As a consequence, the self-diffusion coefficient is always reduced in interacting systems. A nonmonotonic decay is known to exist in active \cite{caraglio2022analytic}, driven \cite{leitmann2018time}, and harmonically trapped systems \cite{franosch2011resonances}, as well as in fluid systems with inertia \cite{jeney2008anisotropic}, but it has never been shown in equilibrium systems. In fact, such a behavior was even shown to be prohibited in overdamped equilibrium systems \cite{feller1971introduction, leitmann2017time, caraglio2022analytic}. The lack of any nonmonotonic features is intuitively expected, for there exists neither inertia nor any internal or external driving which could introduce additional time scales to the system. 


Here we show for the first time that the FACF can be nonmonotonic and even oscillatory in overdamped equilibrium systems. Systems showing this behavior are characterized by probability fluxes, which are perpendicular to density gradients and are referred to as \emph{odd-diffusive} systems~\cite{hargus2021odd}. We further demonstrate that the unusual behavior of the self-diffusion coefficient in these systems, it increases with increasing density~\cite{kalz2022collisions}, is a natural consequence of the nonmonotonicity of the FACF. The transverse response to the perturbation is the fundamental property of \emph{odd} systems which have received much interest lately~\cite{fruchart2023odd}. In addition to odd-diffusive systems, there are odd systems characterized by odd viscosity~\cite{banerjee2017odd,han2021fluctuating,markovich2021odd,zhao2022odd,lier2022lift,hargus2020time}, odd elasticity~\cite{scheibner2020odd,braverman2021topological} and odd viscoelasticity~\cite{banerjee2021active,lier2022passive}. With the advent of experimental odd systems such as spinning biological organisms~\cite{tan2022odd}, chiral fluids~\cite{soni2019odd, vega2022diffusive} and colloidal spinners~\cite{bililign2022motile}, the interest in odd systems has increased rapidly.


The odd-diffusion tensor for a two-dimensional isotropic system can be written as
\begin{equation}
\mathbf{D} = D_0 \left( \mathbf{1} + \kappa \boldsymbol{\varepsilon} \right),
\label{diffusion_tensor}
\end{equation}
where $\mathsf{1}$ is the identity matrix, $\boldsymbol{\varepsilon}$ is the antisymmetric Levi-Civita symbol in two dimensions ($\boldsymbol{\varepsilon}_{xy} = - \boldsymbol{\varepsilon}_{yx} = 1$ and $\boldsymbol{\varepsilon}_{xx} = \boldsymbol{\varepsilon}_{yy} = 0$),
$D_0$ is the diffusivity and $\kappa$ is the odd-diffusion parameter.
A nonzero $\kappa$ results in probability fluxes perpendicular to density gradients. 
Examples of odd-diffusive systems are Brownian particles diffusing under the effect of Lorentz force~\cite{czopnik2001brownian, chun2018emergence,vuijk2019anomalous,abdoli2020nondiffusive,abdoli2020correlations,shinde2022strongly}, and diffusing skyrmions \cite{schutte2014inertia, troncoso2014brownian, wiesendanger2016nanoscale, fert2017magnetic, buttner2018theory, weissenhofer2021skyrmion}, see also the Supplementary Material (SM) \cite{supplemental-material}.
Although these are 
equilibrium odd-diffusive systems, there exist also driven odd-diffusive systems such as active chiral particles (also called circle swimmers)~\cite{kummel2013circular,van2008dynamics,vuijk2022active,muzzeddu2022active} and strongly damped particles subjected to Magnus~\cite{reichhardt2022active} or Coriolis force~\cite{kahlert2012magnetizing}. In contrast to equilibrium systems which are invariant under time-reversal, the odd-diffusive behavior in nonequilibrium systems is a consequence of broken time-reversal and parity symmetries \cite{hargus2020time}.

While an exact calculation of the FACF is a formidable task, near-exact analytical results can be obtained in the dilute limit in which the dynamics are dominated by two-body effects.
To this end, we generalize the first-principles approach developed by Hanna, Hess, and Klein~\cite{hanna1981velocity,hanna1982self} to calculate the FACF in a dilute odd-diffusive system of hard-core interacting particles. We show analytically that odd diffusion qualitatively alters the time correlations: the correlation function becomes negative for finite $\kappa$ indicating the anticorrelated nature of the force experienced by an odd-diffusive particle due to collisions with other particles. Moreover, the correlation function exhibits temporal oscillations for certain values of $\kappa$ it crosses zero twice. We further show that for sufficiently large $\kappa$, the integral of the correlation function becomes negative which gives rise to the increase in the self-diffusion coefficient. Using the Green-Kubo relation, we derive exactly the same expression for the self-diffusion coefficient as in Ref.~\cite{kalz2022collisions} which was obtained using an alternative approach. 
\textit{Theoretical background.} We consider a two-dimensional system of two interacting, odd-diffusive hard disks 
with coordinates $\vec{\mathbf{x}} = (\mathbf{x}_1, \mathbf{x}_2)$. The two-particle conditional probability density function for the particles to evolve from $\vec{\mathbf{x}}^\prime$ at time $t^\prime \leq t$ to $\vec{\mathbf{x}}$ at time $t$, $P = P(\vec{\mathbf{x}}, t|\vec{\mathbf{x}}^\prime, t^\prime)$, satisfies the Fokker-Planck equation 
\begin{align}
\label{N_particle_Smoluchwoski_equation_maintext}
\frac{\partial}{\partial t} P &= \nabla_{1} \cdot \mathbf{D}\left[\nabla_1 + \beta \nabla_1 U(r)\right] P\nonumber \\
& \quad + \nabla_{2} \cdot \mathbf{D}\left[\nabla_2 + \beta \nabla_2 U(r)\right] P,
\end{align}
with the odd-diffusion tensor \eqref{diffusion_tensor} and $\nabla_1, \nabla_2$ as the partial differential operator with respect to the coordinates of particle one and two, respectively. $U(r)$ is the potential energy with $r = |\mathbf{x}_1 - \mathbf{x}_2|$ as the relative distance between the particles and $\beta = 1/k_\mathrm{B} T$, where $k_\mathrm{B}$ is the Boltzmann constant and $T$ is the temperature. We assume hard-core interactions between the two disks of diameter $\sigma$, which can be written as
$U(r) = \small \begin{cases} \textstyle \infty, &r \leq \sigma \\ 0, &r > \sigma\end{cases}$. The analytical solution to the two-particle Fokker-Planck equation was obtained for normal-diffusing particles, i.e. $\mathbf{D} = D_0 \mathbf{1}$~\cite{hanna1981velocity,hanna1982self}. While the hard-core interactions are modeled via Neumann boundary conditions in normal-diffusing systems, they are modeled as oblique boundary conditions in odd-diffusive systems due to the transverse fluxes~\cite{abdoli2020stationary,kalz2022collisions}. This has profound consequences for the solution and therefore for the application of our theory. We solve the two-particle problem \eqref{N_particle_Smoluchwoski_equation_maintext} for odd-diffusive hard disks exactly in the SM~\cite{supplemental-material}. 

\textit{Force autocorrelation tensor.} The force autocorrelation tensor (FACT), which is defined as $\mathsf{C}_F(\tau) =
\left< \mathbf{F}(\tau) \otimes \mathbf{F}(0) \right>$, can be written as~\cite{dhont1996introduction}

\begin{align}  
\mathsf{C}_{F}(\tau) &=
\int  \mathrm{d} \vec{\mathbf{x}}
\int  \mathrm{d} \vec{\mathbf{x}}_0 ~
\mathbf{F}\left( \vec{\mathbf{x}} \right)
\otimes
\mathbf{F}\left( \vec{\mathbf{x}}_0 \right) \nonumber \\ &\quad \times P\left( \vec{\mathbf{x}}, \tau| \vec{\mathbf{x}}_0, 0 \right)
P_\mathrm{eq}\left(  \vec{\mathbf{x}}_0 \right),
\label{intermediate_velocity_correlation_maintext}
\end{align}
for $\tau >0$. Here $\mathbf{F}$ is the interaction force acting on a tagged particle due to other particles, $\langle \cdot\rangle$ denotes an ensemble average with the equilibrium distribution $P_\mathrm{eq}\left(  \vec{\mathbf{x}}_0 \right)$ and the outer product is defined as $[\mathbf{A} \otimes \mathbf{B}]_{\alpha\beta} = A_\alpha B_\beta$. Throughout the paper, time is measured in units of $\tau_0 = \sigma^2/(2D_0)$ which is the characteristic timescale of a particle diffusing over a distance of diameter $\sigma$, i.e. $\tau = t/\tau_0.$
The FACT can be calculated from Eq.~\eqref{intermediate_velocity_correlation_maintext} to first order in the density, 
details of which are shown in SM~\cite{supplemental-material}. 
Similar to the diffusion tensor, the FACT can be split in a diagonal and an antisymmetric off-diagonal part:
\begin{equation}
\mathsf{C}_F(\tau) = 
 C_F^\mathrm{diag}(\tau)  \mathsf{1}
+ 
C_F^\mathrm{off}(\tau) \boldsymbol{\varepsilon},
\end{equation}
for $\tau >0$, where $C_F^\mathrm{diag}(\tau)$ and 
$C_F^\mathrm{off}(\tau)$ are the diagonal and antisymmetric off-diagonal elements of the FACT. In Laplace domain they read
\begin{align}
\label{diagonal_fac_tensor}
\tilde{C}_F^\mathrm{diag}(s) &=\frac{2\ \phi}{\beta^2 D_0}\ \frac{K_1\left[\sqrt{s}K_0 + K_1\right]}{\left[\sqrt{s}K_0 + K_1\right]^2 + \left[\kappa K_1\right]^2}, \\
\label{offdiagonal_fac_tensor}
\tilde{C}_F^\mathrm{off}(s) &=\frac{2\ \phi}{\beta^2 D_0}\ \frac{\kappa\ [K_1]^2}{\left[\sqrt{s}K_0 + K_1\right]^2 + \left[\kappa K_1\right]^2}, 
\end{align}
where $K_n = K_n(\sqrt{s})$ is the modified Bessel function of the second kind  of order $n$, $\phi = \pi\ (N/V)\ (\sigma/2)^2$ 
is the area fraction for $N$ particles of diameter $\sigma$ in an area $V$,
and $\tilde{(\cdot)}$ denotes the Laplace transform with $s$ as the Laplace variable conjugate to $\tau$.
Note that the off-diagonal elements $C^\mathrm{off}_F$ are proportional to the odd-diffusion parameter $\kappa$ and therefore vanish in the case of normal diffusion ($\kappa = 0$).  In this case the FACT reduces to $\mathsf{C}_F(\tau) = C_F^\mathrm{diag}(\tau)\ \mathsf{1} = \frac{\mathsf{1}}{2}\langle \mathbf{F}(\tau) \cdot \mathbf{F}(0)\rangle  $, which is the usual FACF in normal systems. 

\begin{figure*}
\centering 
\begin{subfigure}
\centering 
\includegraphics[width=\columnwidth]{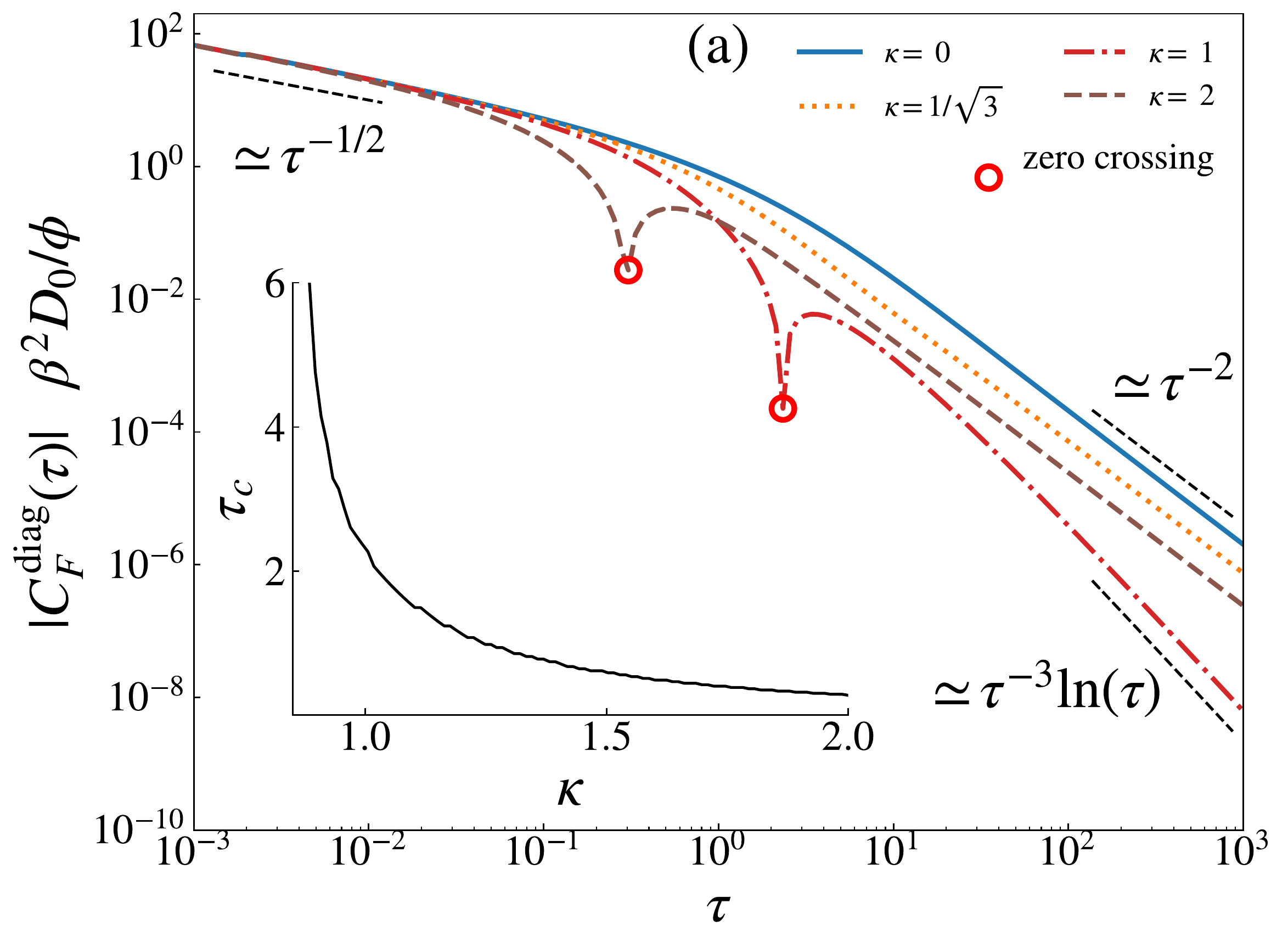}
\end{subfigure}
\begin{subfigure}
\centering
\includegraphics[width=\columnwidth]{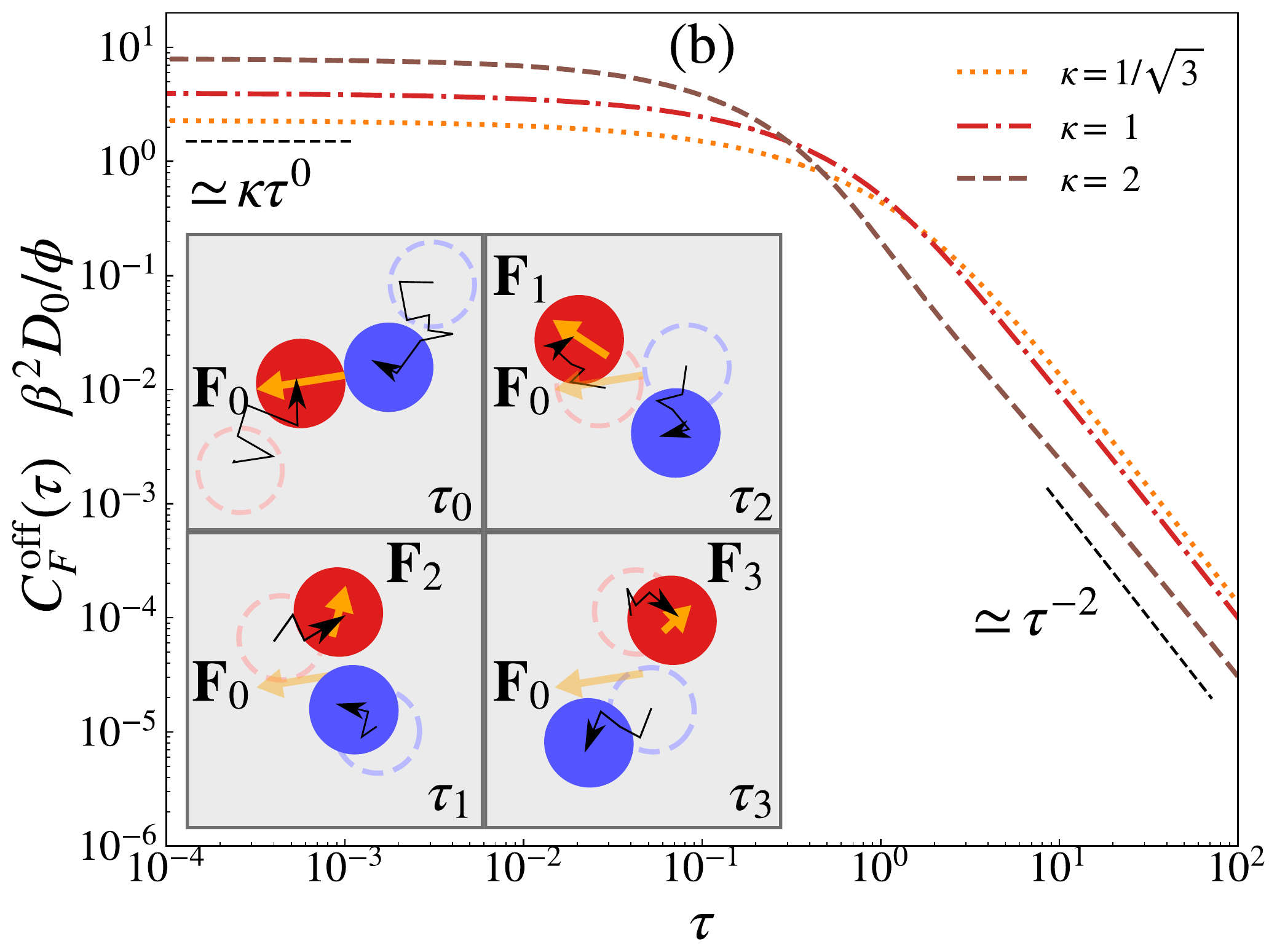}
\end{subfigure}
\caption{Double-logarithmic plot of the diagonal and off-diagonal elements of the force autocorrelation tensor (FACT) of interacting hard disks as a function of reduced time $\tau = t/\tau_0$, where $\tau_0 = \sigma^2/(2D_0)$. (a) The diagonal elements of the FACT $C_F^\mathrm{diag}(\tau)$, corresponding to the force autocorrelation function (FACF), can turn negative. The FACF diverges in the limit $\tau \to 0$ as $C_F^\mathrm{diag}(\tau) \simeq \tau^{-1/2}$. At long times the FACF scales as $C_F^\mathrm{diag}(\tau) \simeq \tau^{-2}$. For $\kappa = 1$ we find an exceptional long-time behavior, where $C_F^\mathrm{diag}(\tau) \simeq \tau^{-3}$. The inset shows the zero-crossing time $\tau_c$ of $C_F^\mathrm{diag}(\tau)$ as a function of $\kappa$, which in the main figure is marked by red circles. The onset of the anticorrelation corresponds to $\kappa > \kappa_{th} \approx 0.88$
(b) The off-diagonal elements of the FACT $C_F^\mathrm{off}(\tau)$ are independent of time in the short-time limit $C_F^\mathrm{off}(\tau) \simeq \kappa \tau^0$ and are directly proportional to $\kappa$. In the long-time limit, they scale similarly to the diagonal elements as $C_F^\mathrm{off }(\tau) \simeq \tau^{-2}$ for all $\kappa$.
The inset in (b) shows typical configurations after a collision of particles, where the orientational change of the force (orange arrow) $\mathbf{F}_i = \mathbf{F}(\tau_i),\ i \in \{0,1,2,3\}$ of the tagged particle (red) is indicated.}
\label{fig:FACF_loglog}
\end{figure*}

The diagonal and off-diagonal elements of the FACT are plotted in Fig.\ref{fig:FACF_loglog} as a function of time. We first consider the behavior of the diagonal elements of the tensor in Fig.\ref{fig:FACF_loglog}(a), which correspond to the usual FACF for odd-diffusive systems. For small values of $\kappa$, the FACF is a positive, monotonically decaying function of time, qualitatively similar to a normal diffusive system. For larger values of $\kappa$, however, a new feature appears in the FACF: it crosses through zero and hence becomes negative, indicating an anticorrelation of the force. 
The time scale of the force reversal on a tracer particle, i.e., when the FACF becomes negative, depends strongly on $\kappa$, as can be seen in the inset of  Fig.\ref{fig:FACF_loglog}(a). There exists a numerically obtained threshold $\kappa_{th} \approx 0.88$ below which the FACF is strictly positive.
The off-diagonal elements of the FACT are shown in Fig.\ref{fig:FACF_loglog}(b). Unlike the diagonal elements, which diverge as $t \to 0$, the off-diagonal elements remain finite. Specifically they remain positive for all $\kappa$ and decay monotonically in time.




It is interesting to investigate the short- and long-time behavior of the elements of the FACT. Using the asymptotic behavior of the modified Bessel functions $K_0$ and $K_1$, see SM~\cite{supplemental-material} 
for details, from Eq.~\eqref{diagonal_fac_tensor} and Eq.~\eqref{offdiagonal_fac_tensor} we have analytical access to the behavior on time scales $t\ll\tau_0$ and $t\gg\tau_0$, i.e. $s \gg 1$ and $s \ll 1$ in the Laplace domain, respectively. 
At short times, the FACF behaves like $C^\mathrm{diag}_F(\tau) \simeq \tau^{-1/2}$, as shown in Fig.\ref{fig:FACF_loglog}(a), and is independent of $\kappa$. Here $\simeq$ is used to denote asymptotic proportionality.
The long-time behavior of the FACF can be obtained from the $s \ll 1$ expansion and behaves asymptotically as
\begin{multline}
\label{c_diag_long_time}
\tilde{C}^\mathrm{diag}_F(s) \sim \frac{2\phi}{\beta^2 D_0} \frac{1}{1 + \kappa^2} \Bigg( 1 +  \frac{1 - \kappa^2}{1 + \kappa^2} \bigg(\gamma - \ln(2)+\frac{\ln(s)}{2}\bigg) s \\+ \frac{1 - 6 \kappa^2 + \kappa^4}{8(\kappa^2 + 1)^3}\ s^2 \ln^2(s) \Bigg),
\end{multline}
for $s\to 0$ and where $\gamma = 0.5772$ is the Euler-Mascheroni constant. For $\kappa = 0$, the asymptotic behavior of $\tilde{C}^\mathrm{diag}_F$ coincides with the form reported for related 2d Lorentz gas systems \cite{franosch2010persistent}. Furthermore, from Eq.~\eqref{c_diag_long_time} it can be seen that the long-time behavior of $C_F^\mathrm{diag}(\tau)$ strongly depends on $\kappa$. The FACF decays as $\tau^{-2}$ for all $\kappa$ except for $\kappa = 1$, at which the leading order contribution vanishes in Eq.~\eqref{c_diag_long_time} and $C_F^\mathrm{diag }(\tau) \simeq \tau^{-3} \ln(\tau)$, as shown in Fig.\ref{fig:FACF_loglog}(a)~\cite{hull1955asymptotic, abramowitz1968handbook}. 
The ordinary algebraic long-time decay $\simeq \tau^{-2}$ ($\kappa \neq 1$) is consistent with the general prediction of a decay $\simeq \tau^{-(d/2 +1)},\ d = 1,2,3,$ for correlation functions in systems, which do not conserve momentum \cite{ernst1971long, van1982transport}. This universal behavior was theoretically and numerically exhaustively demonstrated specifically for the 2d Lorentz gas model \cite{alder1978long, lewis1978evidence,frenkel1987velocity, hofling2007crossover, franosch2010persistent}. In three dimensions, the decay of the correlation functions $\simeq \tau^{-5/2}$ \cite{jacobs1977macromolecular, hanna1981velocity, ackerson1982correlations, felderhof1983cluster, felderhof1983diffusion} could recently be demonstrated computationally \cite{mandal2019persistent}. In contrast, the short-time behavior $\simeq \tau^{-1/2}$ is independent of dimensionality and attributed to the hard interactions between the particles \cite{ hanna1981velocity, ackerson1982correlations}.

The asymptotic short-time behavior of $C_F^\mathrm{off}(\tau)$ turns out to be independent of time but depends linearly on $\kappa$, $C_F^\mathrm{off}(\tau) \simeq \kappa \tau^{0}$, as as can be seen in Fig.\ref{fig:FACF_loglog}(b). Such a scaling of the off-diagonal elements with $\kappa$ at short times has been recently derived by Yasuda et. al in Ref.~\cite{yasuda2022time} for odd Langevin systems.
The authors also pointed out that this could be useful for estimating the odd-diffusion parameter in experiments. The asymptotic long-time behavior of $C_F^\mathrm{off}(\tau)$ shows a monotonic decay in time and also depends on $\kappa$, $C_F^\mathrm{off}(\tau) \simeq \kappa \tau^{-2}/(\kappa^2 + 1)^2$, as can be seen in Fig.\ref{fig:FACF_loglog}(b).

\begin{figure}
\centering 
\includegraphics[width=\columnwidth]{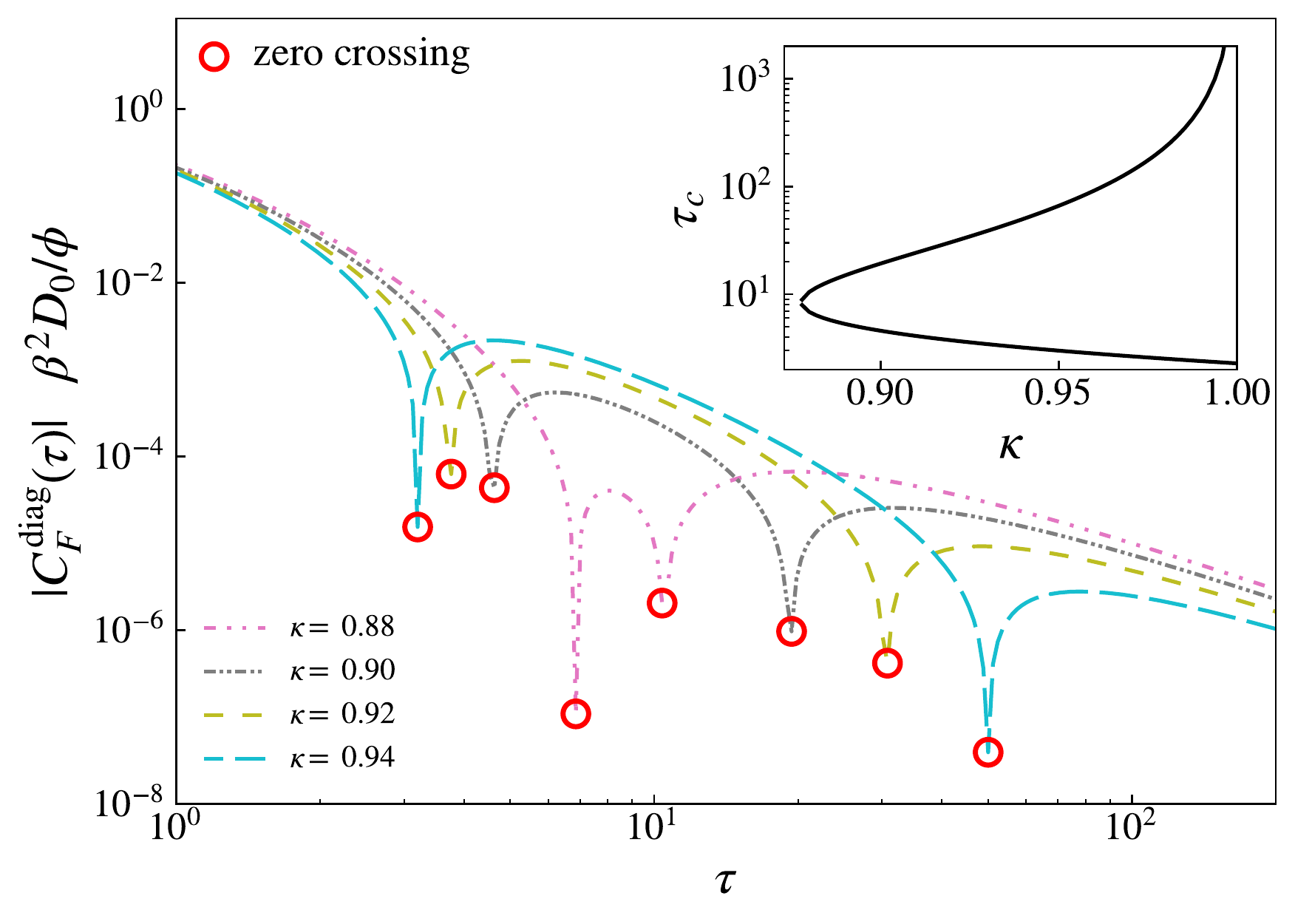}
\caption{Double-logarithmic plot of the absolute value of the diagonal elements of the force autocorrelation tensor $C_F^\mathrm{diag}(\tau)$ of interacting hard disks as a function of reduced time $\tau = t/\tau_0$, where $\tau_0 = \sigma^2/(2D_0)$. Investigating the regime $\kappa \in [0.88, 1.0]$, we find oscillatory behavior of $C_F^\mathrm{diag}(\tau)$. At short times $C_F^\mathrm{diag}(\tau)$ starts as a positive function, turns negative and after a second zero-crossing becomes positive again.  The inset shows the zero-crossing times $\tau_c$ of $C_F^\mathrm{diag}(\tau)$ as a function of $\kappa$ in a linear-logarithmic plot which in the main figure are marked as red circles. The oscillatory behavior starts at $\kappa \geq \kappa_{th} =  0.88$, whereas the second zero-crossing drifts to infinity as $\kappa \to 1$. For $\kappa > 1$, $C_F^\mathrm{diag}(\tau)$ only shows one zero-crossing and remains anti-correlated for the remaining $\tau \to \infty$ (see also inset in Fig.\ref{fig:FACF_loglog} (a)).
}
\label{fig:FACF_ZC}
\end{figure}

In a low-density system, in which only two-body correlations are important, it is quite surprising that the FACF can turn negative, as shown in Fig.\ref{fig:FACF_loglog}. It is even more surprising that there exists a range of $\kappa \in (\kappa_{th}, 1)$ for which the FACF exhibits not one but two zero crossings, as shown in Fig.~\ref{fig:FACF_ZC}. 
It appears that for $\kappa$ slightly larger than $\kappa_{th} \approx  0.88$, which is obtained from numerical inversion of Eq.~\eqref{diagonal_fac_tensor}, the FACF first becomes anticorrelated (first zero crossing) in time before it crosses the time axis again (second and last zero crossing). Here, at long times, the FACF decays to zero from above. We have numerically inverted the Laplace transform over much longer times than shown here and did not find more than two zero crossings. 
This ''temporal oscillation'' in the FACF ceases to exist for $\kappa \geq 1$. For $\kappa > 1$, the asymptotic expansion in Eq.~\eqref{c_diag_long_time}, transformed back into time domain, is strictly negative and therefore the FACF decays to zero from below, i.e. the second zero-crossing vanishes (see also inset in Fig.\ref{fig:FACF_ZC}).

\textit{Green-Kubo relation for the self-diffusion coefficient.} The self-diffusion coefficient $D_s$ can be obtained from the velocity autocorrelation function (VACF) $C_v(\tau) = \langle \mathbf{v}(\tau) \cdot \mathbf{v}(0)\rangle/2$,
where $\mathbf{v}(\tau) = \mathrm{d} \mathbf{x} / \mathrm{d}\tau$ and $\mathbf{x}$ is the position of the tagged particle as the time integral
\begin{equation}
\label{green_kubo_velocity}
D_s = \int_0^\infty\mathrm{d}t\ C_v(\tau), 
\end{equation}
a Green-Kubo relation between an equilibrium autocorrelation function ($C_v(\tau)$) and a transport coefficient ($D_s$)~\cite{green1954markoff}. 



In normal diffusive systems, the VACF is related to the FACF. In contrast, in an odd-diffusive system, the knowledge of the FACF alone is not sufficient to calculate the VACF. This is despite the fact that the system is isotropic. In fact, one requires the entire FACT to calculate the velocity correlation function. We show in SM~\cite{supplemental-material} 
that in odd-diffusive systems, the VACF can be written as 
\begin{equation}
\label{vacf_realted_to_facf}
C_v(\tau) =  D_0 \left(\delta_+(\tau) - D_0 \beta^2 C_F(\tau)  \right), 
\end{equation}
where 
\begin{equation}
\label{mfacf_as_contraction}
C_F(\tau) = \frac{1}{2} \frac{1}{D_0^2} (\mathsf{D}^2)^\mathrm{T} : \mathsf{C}_F(\tau),
\end{equation}
and where the double contraction is defined as $\mathsf{A} : \mathsf{B} = \sum_{\alpha, \beta = 1}^2A_{\alpha\beta} B_{\beta\alpha}$. $\delta_+$ is the one-sided delta distribution, see also the SM~\cite{supplemental-material}. We refer to $C_F(\tau)$ as the \textit{generalized} force autocorrelation function (gFACF) which reads
\begin{equation}
\label{contracted_mfacf_function}
C_F(\tau) = \left(1 - \kappa^2\right)C_F^\mathrm{diag}(\tau) - 2\kappa C_F^\mathrm{off}(\tau).
\end{equation}
For normal diffusive systems (i.e. $\kappa = 0$), $C_F$ reduces to the ordinary FACF. Note that even though the gFACF is diverging for all $\kappa \neq 1$ in $\tau \rightarrow 0$ in the hard-disk system, the function remains integrable. This is of physical significance since the integral of the gFACF captures the effect of collisions on the self-diffusion as we see from the Green-Kubo relation Eq.~\eqref{green_kubo_velocity} together with Eq.~\eqref{vacf_realted_to_facf}.

The self-diffusion coefficient $D_s$ can be obtained from the time integral of Eq.~\eqref{vacf_realted_to_facf} or by using the limit theorem $\int_0^\infty f(t)\ \mathrm{d}t = \lim_{s \to 0} \tilde{f}(s)$ in Eq.~\eqref{c_diag_long_time} for $\tilde{C}^{\mathrm{diag}}_F$ and similarly for $\tilde{C}^{\mathrm{off}}_F$ which yields

\begin{align}
\lim_{s \to 0} \tilde{C}_F^\mathrm{diag}(s) &= 
\frac{1}{\kappa} \lim_{s \to 0} \tilde{C}_F^\mathrm{off}(s) =  \frac{2\ \phi}{\beta^2 D_0}\ \frac{1}{1+ \kappa^2}.
\end{align}

Together with Eq.~\eqref{vacf_realted_to_facf} and Eq.~\eqref{contracted_mfacf_function}, this gives the self-diffusion coefficient in an odd-diffusive system,
\begin{equation}
\label{self_diffusion_result}
D_s = D_0 \left(1 - 2\phi\ \frac{1 - 3\kappa^2}{1 + \kappa^2}\right),
\end{equation}
valid up to first order in area concentration $\phi$ for a system of hard disks.
This result was previously derived by us in Refs.~\cite{kalz2022collisions, kalz2022diffusion} by a different method.

For $\kappa = 0$ the expression for $D_s$ reproduces the known result of normal diffusive systems of hard disks in two dimensions $D_s = D_0 (1 - 2\phi)$~\cite{hanna1982self, ackerson1982correlations}.
The surprising result of $D_s$ in Eq.~\eqref{self_diffusion_result} is that the prefactor of $\phi$ can change sign.
This shows that odd diffusivity ($\kappa>0$) results in a cancellation of the ordinary collision-induced reduction of the self-diffusion.
For $\kappa = \kappa_c = 1/ \sqrt{3}$, up to first order in the area fraction, the effect of the collisions on the self-diffusion vanishes ($D_s = D_0$), meaning that the 
on long time and length scales hard disks appear to diffuse as non-interacting particles.
For $\kappa > \kappa_c$, collisions surprisingly increase the self-diffusion coefficient: the system mixes more efficiently. 

It is natural to ask whether our findings can be extended to three dimensions. However, in three dimensions, odd systems cannot be isotropic because the plane in which the rotation takes place breaks isotropy~\cite{fruchart2023odd,avron1998odd,hargus2021odd}. We investigated the self-diffusion in such a system via Brownian dynamics simulations and found that the in-plane odd diffusivity has no effect on the diffusion along the axes of rotation, which turns out to be exactly the same as that of a normal-diffusive system of hard spheres. The in-plane diffusivity, however, shows the same $\kappa$-dependent behavior as in a two-dimensional odd-diffusive system.

\textit{Discussion.} 
We analytically demonstrated that equilibrium correlation functions can be non-monotonic and even oscillatory in overdamped systems. This finding is at odds with the statement that in an equilibrium system the correlation function and all its derivatives decay monotonically~\cite{leitmann2017time, caraglio2022analytic}.  While the latter holds in systems where the time-evolution is described by a Hermitian Fokker-Planck operator, for odd systems this is not applicable due to their intrinsic antisymmetric off-diagonal elements in the diffusion tensor \eqref{diffusion_tensor}.

Our work shows that rich physics is to be explored in equilibrium, odd-diffusive systems. In normal-diffusive systems, for instance, there exists a crossover between two diffusive regimes: short-time diffusion with diffusivity $D_0$ and long-time diffusion with $D_s < D_0$~\cite{dhont1996introduction}. That the long-time self-diffusion coefficient is smaller than the short-time is indicative of the slowing down of the dynamics of the tracer particle in the crossover. In odd-diffusive systems, in contrast, the dynamics can be enhanced, which is reflected in the anticorrelated force autocorrelations.
The anticorrelation can be physically interpreted in terms of reversal of the force experienced by a tagged particle such that rather than impeding, collisions with other odd-diffusive particles enhance the motion of the tagged particle, see also the inset in Fig. \ref{fig:FACF_loglog}(b). Even though qualitatively this mutual rolling of particles explains the enhancement of self-diffusion with collisions in an odd-diffusive system through the reversal of force~\cite{kalz2022collisions}, a detailed mechanism is still elusive.
To this end, we believe it will be interesting to investigate the structural rearrangements that occur in an odd-diffusive system and contrast them with those in a normal diffusive system. We further expect that the unusual behavior could also have implications for the rheological properties of odd fluids, such as viscosity.
With increasing experimental interest in systems such as spinning biological organisms ~\cite{tan2022odd}, chiral fluids ~\cite{soni2019odd, vega2022diffusive}, and colloidal spinners~\cite{bililign2022motile}, our work will contribute to the broadening interest of the physics community in these systems, especially in the novel and interesting way how interactions modify the particle dynamics here. Furthermore, we believe that our work will stimulate fundamental research on extending statistical physics to the novel case of odd-diffusive systems. Lastly, since exact analytical results are rather rare in interacting systems, our work may serve as a reference to validate approximate theories for dense systems or computer simulations. 



\begin{acknowledgements} \textit{Acknowledgements.} We would like to thank one of the anonymous Reviewers for several suggestions and fruitful comments regarding the analytical
results of the auto-correlation functions discussed in this work. We further thank Felix B{\"u}ttner for illuminating discussions on skyrmionic systems. E. K., R. M., and A. S. acknowledge support by the Deutsche Forschungsgemeinschaft (grants No. ME 1535/16-1 and SH 1275/3-1). J.-U.S. thanks the cluster of excellence ``Physics of Life'' at TU Dresden for support.
\end{acknowledgements}

\end{document}


\title{Supplementary Material: Oscillatory force autocorrelations in equilibrium odd-diffusive systems}

\author{Erik Kalz}
\affiliation{University of Potsdam, Institute of Physics and Astronomy, D-14476 Potsdam, Germany}

\author{Hidde Derk Vuijk}
\affiliation{University of Augsburg, Institute of Physics, D-86159 Augsburg, Germany}

\author{Jens-Uwe Sommer}
\affiliation{Leibniz-Institute for Polymer Research, Institute Theory of Polymers, D-01069 Dresden, Germany}
\affiliation{Technical University of Dresden, Institute for Theoretical Physics, D-01069 Dresden, Germany}
\affiliation{Technical University of Dresden, Cluster of Excellence Physics of Life, D-01069 Dresden, Germany}

\author{Ralf Metzler}
\affiliation{University of Potsdam, Institute of Physics and Astronomy, D-14476 Potsdam, Germany}
\affiliation{Asia Pacific Centre for Theoretical Physics, KR-37673 Pohang, Republic of Korea}

\author{Abhinav Sharma}
\affiliation{University of Augsburg, Institute of Physics, D-86159 Augsburg, Germany}
\affiliation{Leibniz-Institute for Polymer Research, Institute Theory of Polymers, D-01069 Dresden, Germany}

\maketitle
In this Supplementary Material, we provide the analytical background to derive the main expressions of the elements of the force autocorrelation tensor and their relation to the self-diffusion in odd-diffusive systems. The Supplementary Material is organized as follows: in Sections~\ref{brownian_particles_under_Lorentz_force} and~\ref{diffusing_skyrmions} we shortly review the archetypal equilibrium odd-diffusive systems of Brownian particles under Lorentz force and diffusing skyrmions. In Section~\ref{FP_solution} we solve the two-particle Fokker-Planck equation for interacting off-diffusive hard disks in two dimensions. In Section~\ref{appendix_from_vac_to_fac} we establish the connection from the velocity to the force autocorrelation function from first principles in odd-diffusive systems. The characteristic here is that we need full information from the full force autocorrelation tensor to restore the dynamics correctly. In Section~\ref{appendix_evaluation_of_fac_tensor} we explicitly solve for the elements of the force autocorrelation tensor in the Laplace domain and give an asymptotic expansion for short and long time. In Section~\ref{integral_expressions_from_Gradshteyn_Ryzhik} we give the necessary integral expressions, used in the calculations.

\section{Brownian particles under the effect of Lorentz force}
\label{brownian_particles_under_Lorentz_force}
Brownian particles under the effect of Lorentz force can be described by the underdamped Langevin equation \cite{czopnik2001brownian, chun2018emergence, vuijk2019anomalous}
\begin{subequations}
\label{underdamped_langevin_eq}
\begin{align}
    \dot{\mathbf{x}}(t) &= \mathbf{v}(t), \\
    m\dot{\mathbf{v}}(t) &= -\gamma \mathbf{v}(t) + q\mathbf{v}(t) \times \mathbf{B} - \mathbf{f}(\mathbf{x}) + \boldsymbol{\xi}(t), 
\end{align}
\end{subequations}
where $\mathbf{x}(t)$ is the position of the particle and $\mathbf{v}(t)$ its velocity. $\boldsymbol{\xi}(t)$ constitutes a Gaussian white noise, that is uncorrelated among the coordinates, i.e. $\langle\boldsymbol{\xi}(t)\rangle = \mathbf{0}$ and $\langle\boldsymbol{\xi}(t)\boldsymbol{\xi}^\mathrm{T}(t^\prime)\rangle= 2\gamma k_\mathrm{B}T\ \mathbf{1}\delta(t -t^\prime)$. $\langle\ldots\rangle$ denotes an equilibrium average and $(\cdot)^\mathrm{T}$ a matrix transpose. $m, \gamma, q$ are the particle mass, friction, and charge, and $k_\mathrm{B}T$ is the temperature of the solvent in units of the Boltzmann constant. $\mathbf{B}$ is the (constant) external magnetic field, which is taken to point along the $z$-direction. The effect of the Lorentz-force then is only present in the two-dimensional $xy$-plane. It is therefore sufficient to restrict the analysis to two dimensions. $\mathbf{f}(\mathbf{x})$ is an additional external force. Taking the (non-trivial) overdamped limit of the upper Langevin equation \cite{chun2018emergence}, the resulting time-evolution equation for the position reads
\begin{equation}
\label{overdamped_odd_langevin_equation}
    \dot{\mathbf{x}}(t) = -\beta \mathbf{D} \cdot \mathbf{f}(\mathbf{x}) + \boldsymbol{\eta}(t). 
\end{equation}
Here $\beta = 1/k_\mathrm{B}T$. $\boldsymbol{\eta}(t)$ constitutes a Gaussian, but non-white noise accounting for the broken time-reversal symmetry due to the Lorentz force, i.e. $\langle \boldsymbol{\eta }(t)\rangle =0$ and $\langle \boldsymbol{\eta}(t)\boldsymbol{\eta}(t^\prime)\rangle = \mathbf{D} \delta_+(t - t^\prime) + \mathbf{D}^\mathrm{T}\delta_-(t - t^\prime)$. Here Chun et al. \cite{chun2018emergence} used variants of the Dirac delta distribution 
$\delta_\pm(u)$ to account for the broken time-reversal symmetry due to the Lorentz force. They are equal to zero for $u \neq 0$, while $\int_0^\infty\mathrm{d}u\ \delta_+(u) = \int_{-\infty}^0\mathrm{d}u\ \delta_-(u)= 1$ and $\int_0^\infty\mathrm{d}u\ \delta_-(u) = \int_{-\infty}^0\mathrm{d}u\ \delta_+(u)= 0$. The diffusion tensor appearing in Eq.~\eqref{overdamped_odd_langevin_equation} and in the time-correlation of the overdamped noise, $\mathbf{D} = \frac{k_\mathrm{B}T/\gamma}{1 + \kappa^2} (\mathbf{1} + \kappa \boldsymbol{\varepsilon})$, is the characteristic odd-diffusion tensor as we introduce it in Eq.(1) of our main manuscript, with 
$\boldsymbol{\varepsilon}$ as the two-dimensional Levi-Civita symbol. In this specific system, the odd-diffusion parameter is given by $\kappa = qB/\gamma$ and we define  $D_0:= \frac{k_\mathrm{B}T/\gamma}{1 + \kappa^2}$ as the bare diffusion coefficient.

By explicit methods, such as for example the Brinkman's hierarchy \cite{risken1996fokker}, one can derive the Fokker-Planck equation for the probability density function (PDF) 
$p(\mathbf{x}, t)$, corresponding to the overdamped Langevin equation in Eq.~\eqref{overdamped_odd_langevin_equation} as
\begin{equation}
\label{one_body_FP}
    \frac{\partial}{\partial t} p(\mathbf{x},t) = \nabla \cdot \mathbf{D} \left[\nabla - \beta \mathbf{f}(\mathbf{x})\right] p(\mathbf{x},t).
\end{equation}

\section{Diffusing Skyrmions}\label{diffusing_skyrmions}

Skyrmions are topological magnetic configurations, which form whirling, typically circular, textures in a homogeneous magnetic phase. They can be observed in complex crystals and ultrathin metallic films. \cite{wiesendanger2016nanoscale, fert2017magnetic, buttner2018theory}. Assuming an effective two-dimensional rigid body like skyrmion, its movement under the influence of an external force $\mathbf{f}(\mathbf{x})$ is described by the Thiele equation \cite{thiele1973steady},
\begin{equation}
\label{Thiele_equation}
    \mathbf{G} \times \mathbf{v}(t) + \boldsymbol{\gamma}\cdot \mathbf{v}(t) -\mathbf{f}(\mathbf{x}) = 0,
\end{equation}
where $\mathbf{v}(t) = \dot{\mathbf{x}}(t) = (\dot{x}(t), \dot{y}(t))^\mathrm{T}$ is the two-dimensional velocity in Cartesian coordinates, assumed to be constant in the Thiele model. $\mathbf{G} = (0,0,G)^\mathrm{T}$ is the gyrovector, an effective magnetic field pointing perpendicular to the skyrmion-plane, assumed to be the Cartesian $xy$-plane here, and $\boldsymbol{\gamma}$ is the friction tensor. The expressions in the Thiele equation can be made explicit by connecting them to first-principles equations for the time evolution of the spin variables. Considering fluctuations in the local magnetic fields, the governing equations are known as the stochastic Landau-Lifschitz-Gilbert equations \cite{landau1992theory, gilbert2004phenomenological, nowak2007classical}. In terms of the direction of magnetization $\mathbf{s}(\mathbf{x},t)$, the gyromagnetic vector can be derived as $G = \hbar s_0\ \int\mathrm{d}\mathbf{x}\ \mathbf{s}\cdot (\partial \mathbf{s}/\partial x) \times (\partial \mathbf{s}/\partial y)$, also given in terms of the topological charge $Q$ as $G = 4\pi \hbar s_0 Q$, independent on the microscopic details. Here $s_0$ is the local spin density and $\hbar$ the Planck constant divided by $2\pi$. The friction tensor is given by $\boldsymbol{\gamma} = \alpha \mathcal{D}$, where $\alpha$ is the Gilbert damping and $\mathcal{D}$ the so-called dissipation tensor. Due to the assumed rotational symmetry of the skyrmion, it is usually taken to be proportional to the identity and given as $\mathcal{D} = \hbar s_0  \mathbf{1}\  \int\mathrm{d}\mathbf{x}\ [(\partial \mathbf{s}/\partial x)^2 + (\partial \mathbf{s}/\partial y)^2]/2$ \cite{schutte2014inertia}.

At finite temperatures, a skyrmion is coupling to equilibrium electronic and phononic baths. This is expressed via local fluctuations of the magnetic field in the Landau-Lifschitz-Gilbert equation, and as a result, the Thiele equation \eqref{Thiele_equation} is augmented with a random force contribution \cite{schutte2014inertia, troncoso2014brownian}. It was further recently suggested that at high temperatures, a skyrmion also couples to an equilibrium bath of magnons \cite{weissenhofer2021skyrmion}, and as a result, the friction has to be replaced by effective friction $\gamma_\mathrm{eff} = \alpha \mathcal{D} + \eta T$, with the (phenomenological) coupling constant $\eta$. The stochastic Thiele equation finally is form-equivalent to the Langevin equation of motion for Brownian particles under the effect of the Lorentz force \eqref{overdamped_odd_langevin_equation}, which renders diffusing skyrmions obey odd dynamics in the overdamped limit. They are shown to follow diffusive motion with a bare diffusion coefficient $D_0 = k_\mathrm{B}T/(\gamma_\mathrm{eff}(1 + \kappa^2))$ \cite{weissenhofer2021skyrmion, schutte2014inertia, troncoso2014brownian}, where the odd-diffusion parameter is given by $\kappa = G/\gamma_\mathrm{eff}$. The diffusion of skyrmions has recently also been realized experimentally \cite{gruber2023300}.

\section{Solution of the two particle Fokker-Planck equation}
\label{FP_solution}

This section closely follows Ref. \cite{hanna1982self} in its arguments and adapts them to odd-diffusive systems.

\subsection{The two particle problem}
The Fokker-Planck equation for the joint transition PDF of two interacting Brownian particles at positions $(\mathbf{x}_1, \mathbf{x}_2)$ at time 
$t\geq 0$, given that they were at positions $(\mathbf{x}_1^0, \mathbf{x}_2^0) $ at time $t = 0$, $\mathcal{P}(t) \equiv \mathcal{P}(\mathbf{x}_1, \mathbf{x}_2, t| \mathbf{x}_1^0, \mathbf{x}_2^0, 0)$,  reads
\begin{align}
\frac{\partial}{\partial t} \mathcal{P}(t) &= \nabla_1 \cdot \mathbf{D} \left[\nabla_1 + \beta \nabla_1 U(\mathbf{x}_1, \mathbf{x}_2)\right] \mathcal{P}(t) \nonumber\\ 
\label{original_FP_eq}
 &\quad + \nabla_2 \cdot \mathbf{D} \left[\nabla_2 + \beta \nabla_2 U(\mathbf{x}_1, \mathbf{x}_2)\right] \mathcal{P}(t),
\end{align}
where we recall the odd-diffusion tensor $\mathbf{D} = D_0\left( \mathsf{1} + \kappa \boldsymbol{\varepsilon}\right)$ with $D_0$ as the diffusivity and $\kappa$ the odd-diffusion parameter. $\nabla_i$ are the partial derivatives taken with respect to the $i$th particles position coordinate and $U(\mathbf{x}_1, \mathbf{x}_2)$ is the interaction potential between the particles, specifying the additional interaction force of Eq.~\eqref{one_body_FP}, i.e. $\mathbf{f}_i(\mathbf{x}_1, \mathbf{x}_2) = -\nabla_i U(\mathbf{x}_1, \mathbf{x}_2)$. The initial condition on Eq.~\eqref{original_FP_eq} is 
\begin{equation}
\mathcal{P}(\mathbf{x}_1, \mathbf{x}_2, t = 0| \mathbf{x}_1^0, \mathbf{x}_2^0, 0) = \delta (\mathbf{x}_1 - \mathbf{x}_1^0)\ \delta(\mathbf{x}_2 - \mathbf{x}_2^0).
\end{equation}

For hard disks the interaction potential is
\begin{equation}
\label{hard_core_interaction_potential}
U(\mathbf{x}_1, \mathbf{x}_2) = U(r) = \left\{ \begin{array}{ll} \infty, &r \leq \sigma \\ 0, &r > \sigma\end{array} \right.,
\end{equation}
where $r  = |\mathbf{x}_1 - \mathbf{x}_2|$ is the inter-particle distance and $\sigma$ is the particle diameter.


The Fokker-Planck equation Eq.~\eqref{original_FP_eq} can be written in terms of the inner coordinates
$\mathbf{x} = \mathbf{x}_1 - \mathbf{x}_2$ and center-of-mass coordinates $\mathbf{X} = \frac{1}{2} (\mathbf{x}_1 + \mathbf{x}_2)$, with $\nabla_\mathbf{x}$ and $\nabla_\mathbf{X}$ as the corresponding partial derivatives:
\begin{align}
\label{transformed_FP_eq}
\frac{\partial}{\partial t} \mathcal{P}(t) &= \frac{D_0}{2}\nabla_\mathbf{X} \cdot \nabla_\mathbf{X} \mathcal{P}(t) \nonumber \\ & \quad + 2 \nabla_\mathbf{x}\cdot \left[ D_0 \nabla_\mathbf{x} +  \beta\ \mathbf{D} \nabla_\mathbf{x} U(r)\right] \mathcal{P}(t).
\end{align}
The center-of-mass and inner coordinates are decoupled,
which means that the joint PDF can be written as $\mathcal{P}(t) = P(\mathbf{X}, t| \mathbf{X}_0, 0)\ p(\mathbf{x}, t| \mathbf{x}_0, 0)$,
where the coordinates at $t=0$ are now denoted by a subscribt $0$.
Clearly, the diffusion of the center-of-mass coordinate contributes $D_0$ to the self-diffusion constant of the tagged particle.
In contrast, the time evolution of the PDF of the inner coordinate is governed by
\begin{equation}
\label{inner_PF_equation}
\frac{\partial }{\partial t} p(\mathbf{x}, t| \mathbf{x}_0) = \hat{\Omega}\ p(\mathbf{x}, t| \mathbf{x}_0),
\end{equation}
where
\begin{equation}
\label{relative_FP_operator}
\hat{\Omega} = 2 \nabla_\mathbf{x}\cdot \left[ D_0 \nabla_\mathbf{x} +  \beta\ \mathbf{D} (\nabla_\mathbf{x} U(r))\right],
\end{equation}
is the inner Fokker-Planck operator and the initial condition is given by 
\begin{equation}
\label{initial_condition}
p_0 = p(\mathbf{x}, 0| \mathbf{x}_0) = \delta(\mathbf{x} - \mathbf{x}_0)\ \Theta(r_0 - \sigma).
\end{equation}
Here $\Theta(x) $ is the Heaviside function, defined as $\Theta(x) = 1$ for $x> 0$ and zero otherwise, which only allows for valid initial conditions, that is, particles that are separated at $t=0$ by a distance $|\mathbf{x}_0| = r_0 > \sigma$.

\subsection{The problem in relative coordinates}
Equation Eq.~\eqref{inner_PF_equation} can be solved in polar coordinates
$\mathbf{x} = (r, \varphi)$.
In polar coordinates the inner Fokker-Planck operator of Eq.~\eqref{relative_FP_operator} reads
\begin{align}
\label{polar_representation_FP_operator}
\hat{\Omega} &= \frac{2D_0}{r^2} \bigg[ r \frac{\partial}{\partial r}\ r\frac{\partial}{\partial r} + r \frac{\partial}{\partial r}\ r\beta \frac{\partial U(r)}{\partial r} + \frac{\partial^2}{\partial \varphi^2} \nonumber \\ & \quad - \kappa\ r\beta \frac{\partial U(r)}{\partial r} \frac{\partial}{\partial \varphi}\bigg].
\end{align}
Note that for $\kappa =0$, the Fokker-Planck operator in Eq.~\eqref{polar_representation_FP_operator} reduces to that of a normal diffusive, hard-sphere system \cite{hanna1982self}.
The delta distribution in polar coordinates is
\begin{align}
\delta(\mathbf{x} - \mathbf{x}_0) &= \frac{1}{r_0} \delta(r - r_0)\ \delta(\varphi - \varphi_0) \\ 
& = \frac{1}{r_0} \delta(r - r_0) \sum_{n=-\infty}^\infty f_n(\varphi)\ f_{-n}(\varphi_0),
\end{align}
where the angular delta distribution can be expressed in a complete set of orthonormal modes $\{f_n(x)\} = \{1/\sqrt{2\pi}\ \exp(\mathrm{i}nx); n \in \mathbb{Z}\}$. 

As an ansatz for $p(\mathbf{x},t|\mathbf{x}_0)$, we use

\begin{equation}
\label{ansatz_inner_probability}
p(\mathbf{x},t|\mathbf{x}_0) = \Theta(r - \sigma) \sum_{n=-\infty}^\infty R_n(r,t|r_0) f_n(\varphi)\ f_{-n}(\varphi_0),
\end{equation}
where again the Heaviside function accounts for the hard-core exclusion.
Except for the radial functions $R_n$, all contributions in Eq.~\eqref{ansatz_inner_probability} are time independent. Hence the time differentiation in the Fokker-Planck equation Eq.~\eqref{inner_PF_equation} only affects the radial functions $R_n$. Using the polar representation of the operator in Eq.~\eqref{polar_representation_FP_operator} and the orthogonality of $\{f_n\}$ results in an equation for the radial functions $R_n$, which reads
\begin{widetext}
\begin{align}
\label{equation_for_R}
\Theta(r - \sigma)\ \frac{\partial}{\partial t}  R_n(r,t|r_0) = \Theta(r - \sigma)\ \frac{2D_0}{r^2}\left[r \frac{\partial}{\partial r}\ r\frac{\partial}{\partial r} - n^2\right] R_n(r,t|r_0) + \delta(r - \sigma)\ 2D_0\left[\frac{\partial}{\partial r} + \frac{\mathrm{i}n\kappa}{r}\right]R_n(r,t|r_0),
\end{align}
\end{widetext}
where $\exp(-\beta U(r)) = \Theta(r - \sigma)$ was used,
which allows to differentiate the singular interaction potential $\delta(r - \sigma) = -\beta \Theta(r - \sigma)\ \frac{\partial}{\partial r} U(r)$.

In the domain $r > \sigma$, Eq.~\eqref{equation_for_R} is
\begin{equation}
\label{rquation_for_R}
\frac{\partial}{\partial t}  R_n(r,t|r_0) = \frac{2D_0}{r^2}\left[r \frac{\partial}{\partial r}\ r\frac{\partial}{\partial r} - n^2\right] R_n(r,t|r_0).
\end{equation}
Equation Eq.~\eqref{equation_for_R} also gives the no-flux boundary condition for an odd-diffusive system, the so-called oblique boundary condition (see for example \cite{gilbarg2001elliptic})
to be satisfied at $r = \sigma$
\begin{equation}
\label{boundary_condition_r_equal_sigma}
 r \frac{\partial}{\partial r} R_n(r = \sigma,t|r_0)= -\mathrm{i}n\kappa\ R_n(r = \sigma,t|r_0).
\end{equation}
This can be viewed as the extension of the ordinary Neumann no-flux boundary condition to an odd-diffusive system, to which Eq.~\eqref{boundary_condition_r_equal_sigma} reduces for $\kappa =0$. Note that the appearance of the imaginary unit i in this boundary condition can be traced back to the coupling of angular derivative to the radial derivative of the potential in the Fokker-Planck operator in Eq.~\eqref{polar_representation_FP_operator}.
The second boundary condition is given from $\lim_{|\mathbf{x}| \to \infty} p(\mathbf{x}, t| \mathbf{x}_0) = 0$, and reads
\begin{equation}
\label{boundary_condition_r_to_infty}
\lim_{r \to \infty} R_n(r,t|r_0) = 0.
\end{equation}
The initial condition on $p$ in Eq.~\eqref{initial_condition} translates to  
\begin{equation}
\label{initial_condition_R}
R_n(r, t = 0|r_0) = \frac{\delta(r - r_0)}{r_0}.
\end{equation}

\subsection{Solution of the relative problem}
In order to solve for the radial functions $R_n$, they are decomposed into homogeneous solutions $X_n$, with initial condition zero, i.e. $X_n(r, t= 0|r_0) = 0$, and particular solutions $Z_n$, which satisfy the initial condition on $R_n$, i.e. $Z_n(r, t= 0|r_0) = \delta(r - r_0)/r_0$, such that $R_n = X_n + Z_n$. For the particular solutions we make the ansatz 
\begin{equation}
Z_n(r,t|r_0) = \int_0^\infty \mathrm{d}u\ z_n(r,u|r_0)\ \mathrm{e}^{-2D_0 t u^2}.
\end{equation}
Inserting this into Eq.~\eqref{rquation_for_R}, shows that the functions $z_n$ satisfy the Bessel equation with solutions $J_{n}(ur)$ and $Y_n(ur)$ as the Bessel functions of first kind and second kind, respectively. The general solutions for $z_n$ therefore read
\begin{equation}
z_n(r,u|r_0) = A_n (u|r_0)\ J_{n}(ur) + B_n (u|r_0)\ Y_{n}(ur),
\end{equation}
with amplitues $A_n$ and $B_n$ determined by the initial conditions Eq.~\eqref{initial_condition_R}.
Expanding the delta distribution in Bessel functions of the first kind \cite{arfken2012mathematical}
\begin{equation}
\frac{\delta(r - r_0)}{r_0} = \int_0^\infty \mathrm{d}u\ u\ J_n(ur)\ J_n(ur_0),
\end{equation}
which is valid for $n > -1$ and $r, r_0 >0$, shows that $A_n = u\ J_n(ur_0)$ and $B_n = 0$. The particular solutions $Z_n$ therefore read
\begin{equation}
\label{particular_solution}
Z_n(r,t|r_0) = \int_0^\infty \mathrm{d}u\ u\ J_n(ur)\ J_n(ur_0)\ \mathrm{e}^{-2D_0 t u^2}.
\end{equation}

The homogeneous solutions $X_n(r,t|r_0)$ satisfy Eq.~\eqref{rquation_for_R}. Laplace transforming this equation shows that $\tilde{X}_n(r,s|r_0)$, with $s$ as the Laplace variable, satisfy the modified Bessel equation with solutions $I_n(r\sqrt{s/(2D_0)})$ and $K_n(r\sqrt{s/(2D_0)})$ as the modified Bessel functions of first kind and second kind, respectively. Since $I_n$ diverges for $r \to \infty$, they cannot satisfy the boundary condition $\lim_{r \to \infty} \tilde{X}_n = 0$, and therefore they must be omitted. 

Adding the particular solutions $\tilde{Z}_n$ from Eq.~\eqref{particular_solution} and the homogeneous solutions $\tilde{X}_n$, gives the Laplace transform of the radial functions $\tilde{R}_n(r,s|r_0)$:
\begin{align}
\label{solution_for_R}
\tilde{R}_n(r,s|r_0) & = \int_0^\infty \mathrm{d}u\ u\frac{J_n(ur)\ J_n(ur_0)}{s + 2D_0 u^2} \nonumber \\ & \quad+ \tilde{C}_n(s|r_0)\ K_n\left(r\sqrt{\frac{s}{2D_0}}\right),
\end{align}
where the amplitudes $\tilde{C}_n$ are determined by the oblique boundary condition Eq.~\eqref{boundary_condition_r_equal_sigma} at $r = \sigma$,
which gives
\begin{align}
\label{homogeneous_solution_amplitude}
&\tilde{C}_n(s|r_0) = - \int_0^\infty\mathrm{d}u\ u\ \frac{J_n(ur_0)}{s + 2D_0\ u^2}\ \nonumber \\
&\quad \times \frac{u\sigma\ J_n^\prime(u\sigma) + \mathrm{i}n\kappa\ J_n(u\sigma)}{\sigma \sqrt{\frac{s}{2D_0}}\ K_n^\prime\left(\sigma \sqrt{\frac{s}{2D_0}}\right) + \mathrm{i}n\kappa\ K_n\left(\sigma \sqrt{\frac{s}{2D_0}}\right)},
\end{align}
where the primed functions are defined as $f^\prime(c) = \frac{\partial f(x)}{\partial x}\Big|_{x = c}$ for a function $f$. 
According to the ansatz in Eq.~\eqref{ansatz_inner_probability}, the Laplace transform of the conditional PDF $\tilde{p}(\mathbf{x}, s|\mathbf{x}_0)$ reads
\begin{widetext}
\begin{align}
\label{two_body_solution} 
\tilde{p}(\mathbf{x}, s|\mathbf{x}_0) &= \Theta(r - \sigma)\ \sum_{n=-\infty}^\infty \frac{\mathrm{e}^{\mathrm{i}n (\varphi - \varphi_0)}}{2\pi}\ \int_0^\infty\mathrm{d}u\ u\ \frac{J_n(ur_0)}{s + 2D_0\ u^2} \nonumber \\ &\quad \times\left[ J_n(ur) 
 - K_n\left(r \sqrt{\frac{s}{2D_0}}\right)\ \frac{u\sigma\ J_n^\prime(u\sigma) + \mathrm{i}n\kappa\ J_n(u\sigma)}{\sigma \sqrt{\frac{s}{2D_0}}\ K_n^\prime\left(\sigma \sqrt{\frac{s}{2D_0}}\right) + \mathrm{i}n\kappa\ K_n\left(\sigma \sqrt{\frac{s}{2D_0}}\right)}\right].
\end{align}
\end{widetext}

\section{From velocity to force autocorrelation}
\label{appendix_from_vac_to_fac}

This section closely follows Refs. \cite{hanna1981velocity, dhont1996introduction} in their arguments and adapts them to odd-diffusive systems. Here we derive the connection of the velocity to the force autocorrelation for a general system of $N$ interacting particles and later specify it to the low-density limit.

The contribution of the tagged particle to the total concentration fluctuation $c(\mathbf{k},t)$ of the wavevector $\mathbf{k}$ is given as
\begin{equation}
\label{definition_denisty_fluctuation}
c(\mathbf{k},t) = \mathrm{e}^{-\mathrm{i}\mathbf{k} \cdot \mathbf{x}(t)},
\end{equation}
where $\mathbf{x}(t)$ is the position of the tagged particle at time $t$. A Taylor expansion and taking the time-derivative of this equation, gives
\begin{equation}
\label{taylor_expansion_concentration_mode}
\mathbf{k} \cdot \mathbf{v}(t) = \mathrm{i}\ \frac{\partial}{\partial t} c(\mathbf{k},t) + \mathcal{O}(k^2),
\end{equation}
which becomes exact in the limit of $k = |\mathbf{k}| \to 0$. Here $\mathbf{v}(t) = \dot{\mathbf{x}}(t)$. The previous equation can be used to calculate the velocity autocorrelation function $C_v(t,t^\prime)$ of a Brownian particle, which in two dimensions is defined as
\begin{equation}
C_v(t, t^\prime) = \frac{1}{2} \langle \mathbf{v}(t) \cdot \mathbf{v}(t^\prime) \rangle,
\end{equation}
where $\langle \cdot\rangle$ denotes an equilibrium average. The velocity autocorrelation in an isotropic system does not depend on the direction of the wavevector $\mathbf{k}$. Therefore we average over the direction of $\mathbf{k}$ in Eq.~\eqref{taylor_expansion_concentration_mode} to use this equation for the evaluation of $C_v$. For an arbitrary function $f(\mathbf{k})$ this orientational average is given by the integral $1/(2\pi)\ \int\mathrm{d}\hat{\mathbf{k}}\ f(\mathbf{k})$,
where $\hat{\mathbf{k}}$ denotes the unit vector in direction of $\mathbf{k}$. Averaging the product of velocity and wavevector at two different times $t$ and $t^\prime$ yields
\begin{equation}
\frac{1}{2\pi}\int\mathrm{d}\hat{\mathbf{k}}\ \left[\mathbf{k} \cdot \mathbf{v}(t)\ \mathbf{k} \cdot \mathbf{v}(t^\prime)\right] = \frac{k^2}{2} \mathbf{v}(t) \cdot \mathbf{v}(t^\prime).
\end{equation}
Together with $1 /(2\pi)\ \int\mathrm{d}\hat{\mathbf{k}}\ C_v(t,t^\prime) = C_v(t,t^\prime)$, we can relate the velocity autocorrelation to the density fluctuations in the limit of low wavelengths via Eq.~\eqref{taylor_expansion_concentration_mode}
\begin{equation}
\label{relation_VACF_to_density_fluctuations}
C_v(t,t^\prime) = - \lim_{k \to 0} \frac{1}{k^2}\ \frac{\partial}{\partial t}\ \frac{\partial}{\partial t^\prime} \left\langle \frac{1}{2\pi}\int\mathrm{d}\hat{\mathbf{k}}\ c(\mathbf{k}, t)\ c(\mathbf{k},t^\prime) \right\rangle.
\end{equation}

There is no explicit time dependence in the model, therefore the velocity autocorrelation function is a function of $(t-t^\prime)$ only. Using this and that up to lowest order in the wave vector 
$\partial_t c(\mathbf{k},t)  = - \partial_t c(-\mathbf{k},t)$ from Eq.~\eqref{taylor_expansion_concentration_mode},
Eq.~\eqref{relation_VACF_to_density_fluctuations} can be written as
\begin{equation}
\label{vacf_relation_to_c}
C_v(t-t^\prime) = - \lim_{k \to 0} \frac{1}{k^2}\ \frac{\partial^2}{\partial t^2} \left\langle \frac{1}{2\pi} \int\mathrm{d}\hat{\mathbf{k}}\ c(\mathbf{k}, t)\ c(-\mathbf{k},t^\prime) \right\rangle.
\end{equation}

We proceed with evaluating $C_v$ 
by writing the equilibrium average as an integral over the configuration space of the particles $\vec{\mathbf{x}}$ for $i \in \{1, \ldots , N\}$
weighted with the PDF $P(\vec{\mathbf{x}}, t)$:
\begin{equation}
\langle f(\mathbf{x}, t)\rangle = \int \mathrm{d}\vec{\mathbf{x}}\ f(\mathbf{x}, t)\ P(\vec{\mathbf{x}}, t).
\end{equation}
For the case of the velocity autocorrelation function in Eq.~\eqref{vacf_relation_to_c}, this average reads
\begin{align}
\label{vacf_average_integral}
& C_v(t-t^\prime) = - \lim_{k \to 0} \frac{1}{k^2}\ \frac{\partial^2}{\partial t^2} \int \mathrm{d}\vec{\mathbf{x}} \int\mathrm{d} \vec{\mathbf{x}}^\prime\ \nonumber\\ & \quad \times \frac{1}{2\pi} \int\mathrm{d}\hat{\mathbf{k}}\ c(\mathbf{k}, t) c(-\mathbf{k}, t^\prime)\  P(\vec{\mathbf{x}}, t, \vec{\mathbf{x}}^\prime, t^\prime). 
\end{align}

Using the definition of the density fluctuation in Eq.~\eqref{definition_denisty_fluctuation} together with rewriting the joint PDF in terms of the conditional distribution $P(\vec{\mathbf{x}}, t, \vec{\mathbf{x}}^\prime, t^\prime) = P(\vec{\mathbf{x}}, t|\vec{\mathbf{x}}^\prime, t^\prime)\ P(\vec{\mathbf{x}}^\prime, t^\prime)$, we can rewrite Eq.~\eqref{vacf_average_integral} as
\begin{align}
\label{VACF_in_configuration_space}
& C_v(t - t^\prime) = - \lim_{k \to 0} \frac{1}{k^2}\ \frac{\partial^2}{\partial t^2} \int \mathrm{d}\vec{\mathbf{x}} \int\mathrm{d} \vec{\mathbf{x}}^\prime\  \frac{1}{2\pi} \int\mathrm{d}\hat{\mathbf{k}}\ \nonumber \\ & \quad \times \mathrm{e}^{-\mathrm{i}\mathbf{k}\cdot\mathbf{x}_1}\ P(\vec{\mathbf{x}}, t|\vec{\mathbf{x}}^\prime, t^\prime)\  \mathrm{e}^{\mathrm{i}\mathbf{k}\cdot\mathbf{x}_1^\prime}\ P_\mathrm{eq}(\vec{\mathbf{x}}^\prime),
\end{align}
where we have assumed that at time $t^\prime \leq t$ the system was in equilibrium, i.e. $P(\vec{\mathbf{x}}^\prime, t^\prime) = P_\mathrm{eq}(\vec{\mathbf{x}}^\prime) = Z_\mathrm{eq}^{-1}\ \mathrm{e}^{-\beta U_N(\vec{\mathbf{x}})}$, where $U_N$ is the $N$-particle interaction potential and $Z_\mathrm{eq}$ is the partition function. Note that in Eq.~\eqref{VACF_in_configuration_space} we assigned particle one to be the tagged particle. 

To proceed we need the Laplace transformation of
Eq.~\eqref{VACF_in_configuration_space}.
Let $s$ denote the Laplace variable, then the Laplace transformation $\mathcal{L}(\cdot)$, which we also denote with a tilde, of the second time derivative of the conditional PDF results in 
\begin{align}
\label{laplace_transformed_P} 
& \mathcal{L}\left(\frac{\partial^2}{\partial t^2} P(\vec{\mathbf{x}}, t|\vec{\mathbf{x}}^\prime, t^\prime)\right) = -s P(\vec{\mathbf{x}}, t = t^\prime|\vec{\mathbf{x}}^\prime, t^\prime)  \nonumber \\ & \quad + s^2 \tilde{P}(\vec{\mathbf{x}}, s|\vec{\mathbf{x}}^\prime, 0) - \dot{P}(\vec{\mathbf{x}}, t = t^\prime|\vec{\mathbf{x}}^\prime, t^\prime),
\end{align}
where $\dot{P}(c) = \frac{\partial P(t)}{\partial t}|_{t = c}$. The last term is identically zero due to the symmetry in $t$ and $t^\prime$ in the system \cite{hanna1981velocity}. With Eq.~\eqref{laplace_transformed_P} we find
\begin{align} 
& \tilde{C}_v(s) = - \lim_{k \to 0} \frac{1}{k^2} \int \mathrm{d}\vec{\mathbf{x}} \int\mathrm{d} \vec{\mathbf{x}}^\prime\ \frac{1}{2\pi} \int\mathrm{d}\hat{\mathbf{k}}\ \mathrm{e}^{-\mathrm{i}\mathbf{k}\cdot\mathbf{x}_1} \nonumber \\ & \quad \times [ s^2 \tilde{P}(\vec{\mathbf{x}}, s|\vec{\mathbf{x}}^\prime, 0) - s P(\vec{\mathbf{x}},t^\prime|\vec{\mathbf{x}}^\prime, t^\prime) ]\nonumber \\ & \quad \times \mathrm{e}^{\mathrm{i}\mathbf{k}\cdot\mathbf{x}_1^\prime}\ P_\mathrm{eq}(\vec{\mathbf{x}}^\prime).
\label{Laplace_transformed_VACF_in_configuration_space}
\end{align}

The $N$-particle conditional PDF satisfies the Fokker-Planck equation 
\begin{equation}
\label{N_particle_Smoluchwoski_equation}
\frac{\partial}{\partial t} P(\vec{\mathbf{x}}, t|\vec{\mathbf{x}}^\prime, t^\prime) = \hat{\Omega}_N
\ P(\vec{\mathbf{x}}, t|\vec{\mathbf{x}}^\prime, t^\prime),
\end{equation}
where the $N$-particle Fokker-Planck operator $\hat{\Omega}_N$ for an odd-diffusive system is defined as 
\begin{equation}
\label{N_particle_Smoluchowski_operator}
\hat{\Omega}_N = \sum_{i=1}^N \nabla_{i} \cdot \mathbf{D}\left[\nabla_i + \beta \nabla_i U_N(\vec{\mathbf{x}})\right].
\end{equation}
We recall that $\mathbf{D} = D_0 \left(\mathsf{1} + \kappa \boldsymbol{\varepsilon}\right)$ is the odd-diffusion tensor with $D_0$ as the diffusivity and $\kappa$ the odd-diffusion parameter. 
With the initial condition
\begin{equation}
P(\vec{\mathbf{x}}, t=t^\prime|\vec{\mathbf{x}}^\prime, t^\prime) = \prod_{i=1}^N \delta(\mathbf{x}_i - \mathbf{x}_i^\prime),
\end{equation}
the formal solution to Eq.~\eqref{N_particle_Smoluchwoski_equation} can be written as
\begin{equation}
\label{formal_solution_N_particle_problem}
P(\vec{\mathbf{x}}, t|\vec{\mathbf{x}}^\prime, t^\prime) = \mathrm{e}^{\hat{\Omega}_N(t - t^\prime)}\ \prod_{i=1}^N \delta(\mathbf{x}_i - \mathbf{x}_i^\prime). 
\end{equation}
We perform a Laplace transform of the above solution and insert that in Eq.~\eqref{Laplace_transformed_VACF_in_configuration_space} which reads as 
\begin{align}
\label{laplace_transformed_VACF_with_formal_solution} 
& \tilde{C}_v(s) = \lim_{k \to 0} \frac{s}{k^2} \Bigg[ 1 - s \int \mathrm{d}\vec{\mathbf{x}}\ \frac{1}{2\pi} \int\mathrm{d}\hat{\mathbf{k}}\ \mathrm{e}^{-\mathrm{i} \mathbf{k}\cdot \mathbf{x}_1}\nonumber \\& \quad \times (s - \hat{\Omega}_N)^{-1} \mathrm{e}^{\mathrm{i} \mathbf{k}\cdot \mathbf{x}_1}\  P_\mathrm{eq}(\vec{\mathbf{x}})\Bigg].
\end{align} 
Further using the operator identity 
\begin{equation}
\label{operator_rewriting}
(s - \hat{\Omega}_N)^{-1} = \frac{1}{s} \left(\mathsf{1} + \frac{\hat{\Omega}_N}{s} + \frac{\hat{\Omega}_N}{s} \left( s - \hat{\Omega}_N\right)^{-1}\hat{\Omega}_N\right), 
\end{equation}
Eq.~\eqref{laplace_transformed_VACF_with_formal_solution} can be rewritten as
\begin{align}
\label{intermediated_laplace_transofmed_VACF}
&\tilde{C}_v(s) = - \lim_{k \to 0} \frac{1}{k^2} \int\mathrm{d}\vec{\mathbf{x}}\ \frac{1}{2\pi} \int\mathrm{d}\hat{\mathbf{k}} \nonumber \\ & \quad \times \bigg[\mathrm{e}^{-\mathrm{i} \mathbf{k}\cdot \mathbf{x}_1}\ \hat{\Omega}_N\ \mathrm{e}^{\mathrm{i} \mathbf{k}\cdot \mathbf{x}_1}\ P_\mathrm{eq}(\vec{\mathbf{x}}) \nonumber \\
& \quad + \mathrm{e}^{-\mathrm{i} \mathbf{k}\cdot \mathbf{x}_1}\ \hat{\Omega}_N (s - \hat{\Omega}_N)^{-1} \hat{\Omega}_N\ \mathrm{e}^{\mathrm{i} \mathbf{k}\cdot \mathbf{x}_1}\  P_\mathrm{eq}(\vec{\mathbf{x}})\bigg].
\end{align}
To evaluate Eq.~\eqref{intermediated_laplace_transofmed_VACF} further, we take advantage of the adjoint of the Fokker-Planck operaor
\begin{equation}
\label{adjoint_operator}
\hat{\Omega}_N^\dagger = \sum_{i=1}^N \left[ \nabla_i - \beta\nabla_i U_N(\vec{\mathbf{x}})\right] \cdot \mathbf{D}^\mathrm{T}\nabla_i, 
\end{equation}
where $(\cdot)^\mathrm{T}$ denotes a matrix transpose, which results in
\begin{align}
\label{inbetween_result_2_laplace_transformed_VACF}
& \tilde{C}_v(s) = \hat{\mathbf{k}} \cdot \mathbf{D} \hat{\mathbf{k}} - \beta^2 \int \mathrm{d}\vec{\mathbf{x}}\ \frac{1}{2\pi} \int\mathrm{d}\hat{\mathbf{k}}\ \left[\hat{\mathbf{k}} \cdot \mathbf{D}\nabla_1 U_N(\vec{\mathbf{x}})\right]\nonumber\\ &\quad \times (s - \hat{\Omega}_N)^{-1}\ \left[\hat{\mathbf{k}} \cdot \mathbf{D}^\mathrm{T}\nabla_1 U_N(\vec{\mathbf{x}})\right] P_\mathrm{eq}(\vec{\mathbf{x}}).
\end{align}
Note the different uses of the $\hat{\ }$-symbol here: once to indicate the unit vector $\hat{\mathbf{k}}$ and once to indicate the operator $\hat{\Omega}_N$. 

Rearranging the terms in Eq.~\eqref{inbetween_result_2_laplace_transformed_VACF} and making use of the identity, valid in two dimensions, 
\begin{equation}
\frac{1}{2\pi} \int\mathrm{d} \hat{\mathbf{k}}\ [\hat{\mathbf{k}} \otimes \hat{\mathbf{k}}] = \frac{1}{2}\ \mathsf{1},
\end{equation}
for an outer product $\hat{\mathbf{k}} \otimes \hat{\mathbf{k}}$ of two unit vectors in $k$-direction, we arrive at the expression 
\begin{widetext}
\begin{align}
\tilde{C}_v(s) &= D_0 - \frac{\beta^2}{2} \left(\mathbf{D}^2\right)^\mathrm{T} : \int \mathrm{d}\vec{\mathbf{x}}\ \left[\left( (s - \hat{\Omega}_N)^{-1}\right)^\dagger \nabla_1 U_N(\vec{\mathbf{x}})\right] \otimes \big[\nabla_1 U_N(\vec{\mathbf{x}})\big] P_\mathrm{eq}(\vec{\mathbf{x}}),
\end{align}
where the double dot denotes the contraction of two tensors and is defined as $\mathsf{A} : \mathsf{B} \equiv A_{\alpha\beta}B_{\beta\alpha}$.

Reintroducing the average on the initial coordinate, we can recognize the Laplace transformed conditional PDF $\tilde{P}(\vec{\mathbf{x}},s|\vec{\mathbf{x}}_i^0,0)$ of Eq.~\eqref{formal_solution_N_particle_problem} to appear

\begin{align}
\label{intermediate_velocity_correlation}
\tilde{C}_v(s) &= D_0 - \frac{\beta^2}{2} \left(\mathbf{D}^2\right)^\mathrm{T} : \int \mathrm{d}\vec{\mathbf{x}}\ \int \mathrm{d}\vec{\mathbf{x}}_i^0\ \left[\nabla_1 U_N(\vec{\mathbf{x}}) \right] \otimes \bigg[\underbrace{(s - \hat{\Omega}_N)^{-1} \prod_{i =1}^N \delta(\mathbf{x}_i - \mathbf{x}_i^0)}_{=\tilde{P}(\vec{\mathbf{x}},s|\vec{\mathbf{x}}_i^0,0)} \nabla_1 U_N(\vec{\mathbf{x}}_i^0)\bigg] P_\mathrm{eq}(\vec{\mathbf{x}}_i^0),
\end{align}
\end{widetext}
where we set $t^\prime = 0$ and accordingly $\vec{\mathbf{x}}^\prime = \vec{\mathbf{x}}_i^0$.
The force on the tagged particle is 
$\mathbf{F}(\vec{\mathbf{x}}) = -\nabla_1 U_N\left(\vec{\mathbf{x}}\right)$, similar for the initial condition.
With this, $\tilde{C}_v(s)$ becomes
\begin{equation}
\label{lapalce_inverse_vacf_facf_relation}
\tilde{C}_v(s) = D_0 - \frac{\beta^2}{2} \left( \mathbf{D}^2\right)^\mathrm{T} :
\tilde{\mathsf{C}}_{F}(s),
\end{equation}
where the Laplace transform of the FACT is defined as
\begin{align}  
\tilde{\mathsf{C}}_{F}(s) &=
\int  \mathrm{d} \vec{\mathbf{x}}_i
\int  \mathrm{d} \vec{\mathbf{x}}_i^0 ~
\mathbf{F}\left( \vec{\mathbf{x}}_i \right)
\otimes
\mathbf{F}\left( \vec{\mathbf{x}}_i^0 \right) \nonumber \\ &\quad \times \tilde{P}\left( \vec{\mathbf{x}}_i, s| \vec{\mathbf{x}}_i^0, 0 \right)
P_\mathrm{eq}\left(  \vec{\mathbf{x}}_i^0 \right).
\label{intermediate_velocity_correlation2}
\end{align}
The inverse Laplace transform of Eq.~\eqref{lapalce_inverse_vacf_facf_relation} shows that the velocity and force autocorrelation in an odd-diffusive system thus are related by the generalized relation
\begin{align}
\label{VACF_and_FACF}
C_v(t) &= D_0\left( \delta_+(t) - D_0\beta^2\ \frac{1}{2 D_0^2} \left(\mathbf{D}^2\right)^\mathrm{T} : \mathsf{C}_F(t) \right) \\
&= D_0\left( \delta_+(t) - D_0\beta^2\ C_F(t)\right),
\end{align}
if we define 
\begin{align}
C_F(t) &= \frac{1}{2 D_0^2} \left(\mathbf{D}^2\right)^\mathrm{T} : \mathsf{C}_F(t)\\
\label{contracted_FACF_with_factors}
&=(1 - \kappa^2)\ C^{\mathrm{diag}}_F(t) - 2\kappa\ C^{\mathrm{off}}_F(t)
\end{align}
as the \textit{generalized} force autocorrelation function (gFACF). $\delta_+(t)$ denotes the one-sided delta distribution, for a definition see Section~\ref{brownian_particles_under_Lorentz_force}. Here $C^{\mathrm{diag}}_F(t)$ and $C^{\mathrm{off}}_F(t)$ are the diagonal and antisymmetric off-diagonal elements of the FACT, i.e. $\mathsf{C}_F(t) = C^{\mathrm{diag}}_F(t) \mathsf{1} + C^{\mathrm{off}}_F(t) \boldsymbol{\varepsilon}$. For normal systems ($\kappa = 0$) $C_F(t) = C^{\mathrm{diag}}_F(t)$ reduces to the ordinary FACF.

An analytic calculation of the FACT is only possible in the dilute limit. Therefore we restrict the following calculation to first order in the area fraction. Furthermore, the potential energy is assumed to be the sum of pair potentials
\begin{equation}
U_N(\vec{\mathbf{x}}) = \frac{1}{2} \sum_{i,j=1}^N U(r_{ij}),
\end{equation}
where $r_{ij} = |\mathbf{x}_{ij}| = |\mathbf{x}_i - \mathbf{x}_j|$ is the distance between particle $i$ and particle $j$.
In the dilute limit only two-body correlations are important,
which means that the correlations between the untagged particles can be ignored.
Therefore, the equilibrium and the conditional PDF are approximated as
\begin{align}
\label{approximated_equilibrium_density}
P_\mathrm{eq}(\vec{\mathbf{x}}_i^0,0) = \frac{1}{V} \prod_{i=2}^N \frac{\Theta(r_{i1}^0 - \sigma)}{V},\\
\label{approximated_conditional_density}
\tilde{P}(\vec{\mathbf{x}},s|\vec{\mathbf{x}}_i^0,0) = \frac{1}{V} \prod_{i=2}^N \tilde{p}(\mathbf{x}_{1i}, s|\mathbf{x}_{1i}^0),
\end{align}
where $\tilde{p}(\mathbf{x}_{1i}, s|\mathbf{x}_{1i}^0) = \tilde{p}(\mathbf{x}_{1i}, s|\mathbf{x}_{1i}^0, 0)$ is the Laplace transform of the conditional two-particle PDF for particle one and particle $i$ and $ \Theta(x)$ is the Heaviside function, defined as $\Theta(x) = 1$ for $x > 0$ and zero otherwise. 
The two-particle PDF is the solution to Eq.~\eqref{N_particle_Smoluchwoski_equation} with $N=2$. This two-particle problem is solved in Appendix~\ref{FP_solution}. Note the factor of $1/V$ in Eq.~\eqref{approximated_equilibrium_density} and Eq.~\eqref{approximated_conditional_density}, which comes from the translational invariance of particle one.

Considering all approximations, all but one coordinate can be integrated out in Eq.~\eqref{intermediate_velocity_correlation2}, which results in
\begin{align}
\label{FACF_intermediate_result_2} 
\tilde{\mathsf{C}}_F(s) &= \frac{N}{V} \int\mathrm{d}\mathbf{x}\int\mathrm{d}\mathbf{x}_0\ \mathbf{F}(\mathbf{x}) \otimes \mathbf{F}(\mathbf{x}_0)\nonumber \\ & \quad \times \tilde{p}(\mathbf{x}, s|\mathbf{x}_0)\ \Theta(r_0 - \sigma).
\end{align}
Note that we renamed $\mathbf{x}_{12} = \mathbf{x}$, $\mathbf{x}_{12}^0 = \mathbf{x}_0$ and $r_0 = | \mathbf{x}_{12}^0|$ for the initial position,
and denoted the interaction forces by $\mathbf{F}(\mathbf{x}) = -\nabla_{\mathbf{x}_{12}} U(r_{12})$ and $\mathbf{F}(\mathbf{x}_0) = -\nabla_{\mathbf{x}_{12}^0} U(r_{12}^0)$. 
After an inverse Laplace transformation Eq.~\eqref{FACF_intermediate_result_2} becomes the FACT in the time domain
\begin{align}
\label{final_FACT}
\mathsf{C}_F(t) &= \frac{N}{V}\int\mathrm{d}\mathbf{x}\int\mathrm{d}\mathbf{x}_0\ \mathbf{F}(\mathbf{x}) \otimes \mathbf{F}(\mathbf{x}_0)\nonumber \\ & \quad \times p(\mathbf{x}, t|\mathbf{x}_0)\ \Theta(r_0- \sigma).
\end{align}
For an evaluation of $\tilde{\mathsf{C}}_F(s)$ in the dilute limit, see Appendix~\ref{appendix_evaluation_of_fac_tensor}.

\begin{figure}
\centering 
\includegraphics[width=\columnwidth]{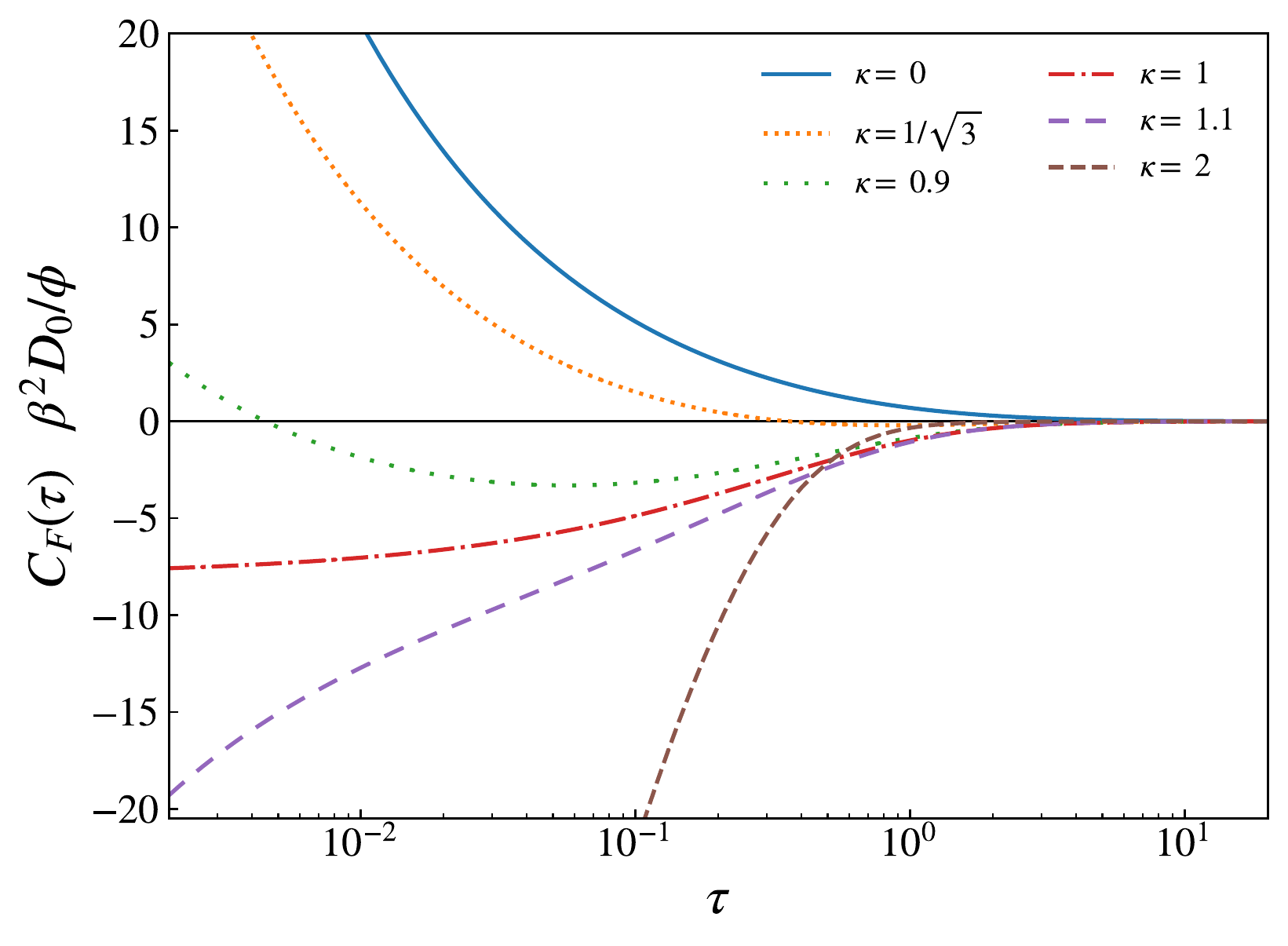}
\caption{The generalized force autocorrelation function (gFACF) $C_F(\tau)$ (Eq.~\eqref{contracted_FACF_with_factors}) for interacting hard disks as a function of reduced time $\tau = t/\tau_0$, where $\tau_0 = \sigma^2/(2D_0)$. For $\kappa > 0$ there exists a range of $\tau$ for which $C_F(\tau)$ is negative.
In the case $\kappa = 1$ the gFACF remains negative and finite for all times.
For $\kappa > 1$ the gFACF is negative for all times and diverges for $t \rightarrow 0$. 
}
\label{fig:gFACF}
\end{figure}

For an explicit evaluation of the gFACF (Eq.~\eqref{contracted_FACF_with_factors}) we also have to use the low-density results of Appendix~\ref{appendix_evaluation_of_fac_tensor}. We show the gFACF for different values of $\kappa$ in Fig.~\ref{fig:gFACF}.
For $0 \leq \kappa < 1$ it diverges to positive infinity as $t\rightarrow 0$,
as expected for the FACF of hard disks \cite{hanna1981velocity}.
For nonzero $\kappa$ the gFACF is negative for a range of time lags.

For $\kappa=1$ the contribution to the gFACF from the diagonal elements of the FACT vanishes,
and therefore the gFACF remains finite as $t \rightarrow 0$.
For $\kappa > 1$ the gFACF diverges to negative infinity as $t \rightarrow 0$. Note that even though the gFACF is diverging for all $\kappa \neq 1$ in $\tau \rightarrow 0$, the function remains integrable. This is of physical significance since the integral of the gFACF captures the effect of collisions on self-diffusion.

\section{Explicit Evaluation of the force autocorrelation tensor}
\label{appendix_evaluation_of_fac_tensor}

The FACT in the Laplace domain Eq.~\eqref{FACF_intermediate_result_2} can be written as
\begin{align}
\label{general_hs_fac_tensor}
\tilde{\mathsf{C}}_F(s) &= \frac{N}{V} \int\mathrm{d}\mathbf{x} \int\mathrm{d}\mathbf{x}_0\  \nabla_\mathbf{x} U(r) \otimes \nabla_{\mathbf{x}_0} U(r_0)\nonumber \\ & \quad \times \tilde{p}(\mathbf{x}, s|\mathbf{x}_0)\ \Theta(r_0 - \sigma), 
\end{align}
where $-\nabla_\mathbf{x} U(r) =\mathbf{F}(\mathbf{x})$ denotes the inter-particle interaction forces and $r = |\mathbf{x}|$, similar for the initial condition. 
The conditional PDF $\tilde{p}(\mathbf{x}, s| \mathbf{x}_0)$ is the Laplace transform of Eq.~\eqref{ansatz_inner_probability}:
\begin{equation}
\label{inner_probaility_for_fac_tensor}
\tilde{p}(\mathbf{x}, s|\mathbf{x}_0) = \Theta(r - \sigma)\ \sum_{n=-\infty}^\infty \tilde{R}_n(r,s|r_0)\ f_n(\varphi)\ f_{-n}(\varphi_0) 
\end{equation}

Using the identity $-\beta U(r) = \ln \exp(-\beta U(r))$,
the singular term $\nabla_\mathbf{x} U(r)$ together with the Heaviside function $\Theta(r-\sigma)$ in $\tilde{p}(\mathbf{x},s|\mathbf{x}_0)$ can be rewritten as
\begin{align}
-\beta \ \Theta(r - \sigma)\ \nabla_\mathbf{x} U(r) &= \Theta(r - \sigma)\ \nabla_\mathbf{x} \ln \Theta(r - \sigma) \\
&= \delta(r - \sigma)\ \hat{\mathbf{x}},
\end{align}
where $\hat{\mathbf{x}}$ represents the unit vector.
The other singular term $\nabla_{\mathbf{x}_0} U(r_0)$ together with $\Theta(r_0-\sigma)$ in Eq.~\eqref{general_hs_fac_tensor}
 can be rewritten in the same way.
This results in
\begin{align}
\label{hs_fac_tensor}
\tilde{\mathsf{C}}_F(s) &= \beta^{-2} \frac{N}{V} \int\mathrm{d}\mathbf{x} \int\mathrm{d}\mathbf{x}_0\ \delta(r - \sigma)\ \delta(r_0 - \sigma)\nonumber \\ & \quad \times  \tilde{p}(\mathbf{x}, s|\mathbf{x}_0)\ \hat{\mathbf{x}} \otimes \hat{\mathbf{x}}_0.
\end{align}

The outer product of the unit vectors $\hat{\mathbf{x}} \otimes \hat{\mathbf{x}}_0$ in Eq.~\eqref{hs_fac_tensor} in polar coordinates is
\begin{equation}
\hat{\mathbf{x}} \otimes \hat{\mathbf{x}}_0 = \left(\begin{array}{*{20}{c}}
\cos(\varphi) \cos(\varphi_0) & \cos(\varphi) \sin(\varphi_0)\\ 
\sin(\varphi) \cos(\varphi_0) & \sin(\varphi) \sin(\varphi_0)
\end{array}\right),
\end{equation}
which can be expressed in terms of the angular functions $\{f_n(x)\} = \{1/ \sqrt{2\pi}\ \exp(\mathrm{i}nx); n \in \mathbb{Z}\}$ as
\begin{align}
\cos(x) &=  \frac{\sqrt{2\pi}}{2} \left(f_1(x) + f_{-1}(x)\right),\\
\sin(x) &= \mathrm{i} \frac{\sqrt{2\pi}}{2} \left(f_{-1}(x) - f_{1}(x)\right).
\end{align}
Using $\mathrm{d}\mathbf{x} = \mathrm{d}r\ \mathrm{d}\varphi\ r$,  $\mathrm{d}\mathbf{x}_0 = \mathrm{d}r_0\ \mathrm{d}\varphi_0\ r_0$ and the orthogonality $\int\mathrm{d}x\ f_n(x)\ f_{m}(x) = \delta_{n, -m}$, where $\delta_{n,m} = 1$ for $n=m$ and zero otherwise
is the Kronecker delta, Eq.~\eqref{hs_fac_tensor} becomes
\begin{widetext}
\begin{align}
\label{facf_tensor_before_radial_integral} 
 \tilde{\mathsf{C}}_F(s) &= \beta^{-2} \pi \frac{N}{V} \sigma^2 \int \mathrm{d}r \int \mathrm{d}r_0\ 
\delta(r - \sigma)\ \delta(r_0 - \sigma) \left[ \left(\tilde{R}_1(r,s|r_0) + \tilde{R}_{-1}(r,s|r_0) \right)
 \mathsf{1}
 - i \left( \tilde{R}_1(r,s|r_0) - \tilde{R}_{-1}(r,s|r_0) \right) \boldsymbol{\varepsilon} \right] \\
 \label{facf_tensor_after_radial_integral}  
  &= 4 \beta^{-2} \phi \left[ \left( \tilde{R}_1(\sigma,s|\sigma) + \tilde{R}_{-1}(\sigma,s|\sigma) \right)
 \mathsf{1}
 - i \left( \tilde{R}_1(\sigma,s|\sigma) - \tilde{R}_{-1}(\sigma,s|\sigma)\right) \boldsymbol{\varepsilon}\right],
\end{align}
\end{widetext}
where $\phi = \pi\ (N/V)\ (\sigma/2)^2$ is the area fraction in two dimensions.

The functions $\tilde{R}_n$ are given in Eq.~\eqref{solution_for_R}.
For $n=\pm 1$ they are
\begin{align}
\label{appendix_R_pm_1}
& \tilde{R}_{\pm 1}(\sigma, s| \sigma) = \frac{1}{2D_0} \int_0^\infty \mathrm{d}x\ \frac{x\ J_{\pm 1}(x)}{x^2 + (s \tau_0)/\sigma^2} \Bigg[J_{\pm 1}(x) \nonumber \\
&\quad - K_{\pm 1}(\sqrt{\tau_0 s})\ \frac{x\ J_{\pm 1}^\prime(x) \pm \mathrm{i}\kappa\ J_{\pm 1}(x)}{\sqrt{\tau_0 s}\ K_{\pm 1}^\prime(\sqrt{\tau_0 s}) \pm \mathrm{i}\kappa\ K_{\pm 1}(\sqrt{\tau_0 s})}\Bigg],
\end{align}
where $\tau_0 = \sigma^2 / (2D_0)$ is the characteristic timescale of a particle diffusing over a distance of diameter $\sigma$.

\textbf{The case of $n = 1$}. We have to find expressions for $x J_1^\prime(x)$ and $b K_1^\prime(b)$, where $b = \sqrt{\tau_0 s}$. The formula for the $k$th derivative of a Bessel function of order $n$ (see \cite{abramowitz1968handbook}, 9.1.30) is given as
\begin{equation}
\label{appendix_differentiation_bessel_function}
\left(\frac{1}{x}\ \frac{\mathrm{d}}{\mathrm{d}x}\right)^k \left( x^n\ J_n(x)\right) = x^{n-k}\ J_{n-k}(x). 
\end{equation}
Therefore $x J_1^\prime(x) = x J_0(x) - J_1(x)$. Similarly, the general formula the $k$th derivative of a modified Bessel function of order $n$ (see \cite{abramowitz1968handbook}, 9.6.28) is given as
\begin{equation}
\label{appendix_differentiation_modified_bessel_function}
\left(\frac{1}{x}\ \frac{\mathrm{d}}{\mathrm{d}x}\right)^k \left( x^n\ \mathrm{e}^{n\mathrm{i}\pi}\ K_n(x)\right) = x^{n-k}\ \mathrm{e}^{(n-k)\mathrm{i}\pi}\ K_{n-k}(x). 
\end{equation}
Therefore $b K_1^\prime(b) = -b K_0(b) - K_1(b)$. According to Eq.~\eqref{appendix_R_pm_1} we find $\tilde{R}_1$ to be
\begin{align}
\label{R_1_integral_expresion}
& \tilde{R}_{1}(\sigma, s| \sigma) = \frac{1}{2D_0} \int_0^\infty \mathrm{d}x\ \frac{x J_{1}(x)}{x^2 + (s \tau_0)/\sigma^2} \Bigg[J_{1}(x) \nonumber \\ &\quad + K_{1}(\sqrt{\tau_0 s})\ \frac{x J_{0}(x) - J_{1}(x)\ (1 - \mathrm{i}\kappa) }{\sqrt{\tau_0 s} K_{0}(\sqrt{\tau_0 s}) + K_{1}(\sqrt{\tau_0 s})(1 - \mathrm{i}\kappa)}\Bigg].
\end{align}
\textbf{The case of $n = -1$}. Here we have to find expressions for $x\ J_{-1}^\prime(x)$ and $b\ K_{-1}^\prime(b)$, where $b = \sqrt{\tau_0 s}$. Using the same rules for differentiation, Eq.~\eqref{appendix_differentiation_bessel_function} for the Bessel functions and Eq.~\eqref{appendix_differentiation_modified_bessel_function} for the modified Bessel functions, together with $J_{-n}(x) = (-1)^n\ J_n(x)$ (see \cite{abramowitz1968handbook}, 9.1.5) and $K_{-n}(x) = K_n(x)$ (see \cite{abramowitz1968handbook}, 9.6.6), we find that $x J_{-1}^\prime(x) = -x J_0(x) + J_1(x)$ and that $b K_{-1}^\prime(b) = -b K_0(b) - K_1(b)$. 
Therefore, $\tilde{R}_{-1}$ is given as
\begin{align}
\label{R_minus1_integral_expresion}
& \tilde{R}_{-1}(\sigma, s| \sigma) = \frac{1}{2D_0} \int_0^\infty \mathrm{d}x\ \frac{x\ J_{1}(x)}{x^2 + (s \tau_0)/\sigma^2} \Bigg[J_{1}(x) \nonumber \\ & \quad + K_{1}(\sqrt{\tau_0 s})\ \frac{x J_{0}(x) - J_{1}(x)\ (1 + \mathrm{i}\kappa) }{\sqrt{\tau_0 s} K_{0}(\sqrt{\tau_0 s}) + K_{1}(\sqrt{\tau_0 s}) (1 + \mathrm{i}\kappa)}\Bigg].
\end{align}

\begin{figure*}
\centering 
\begin{subfigure}
\centering 
\includegraphics[width=\columnwidth]{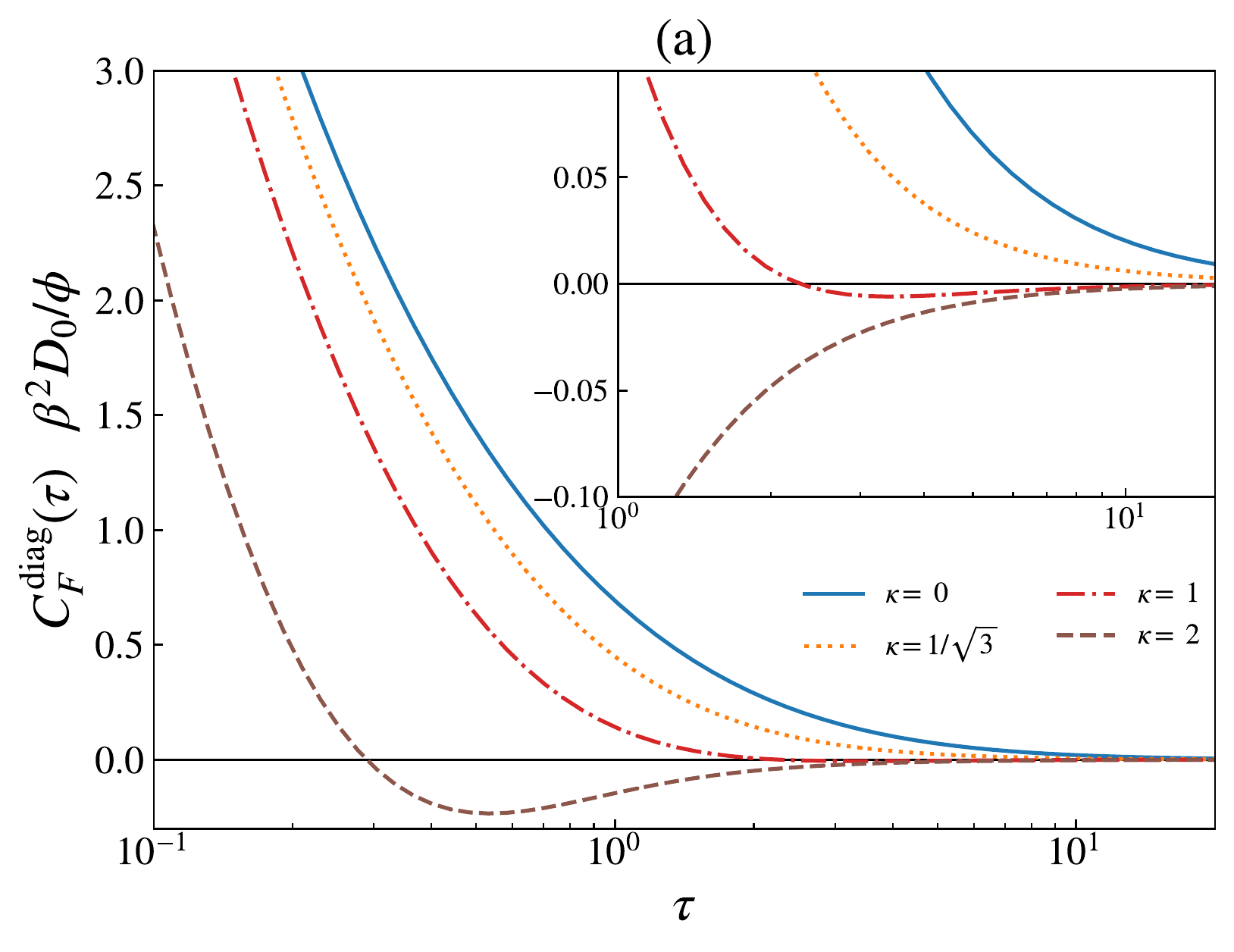}
\end{subfigure}
\begin{subfigure}
\centering 
\includegraphics[width=\columnwidth]{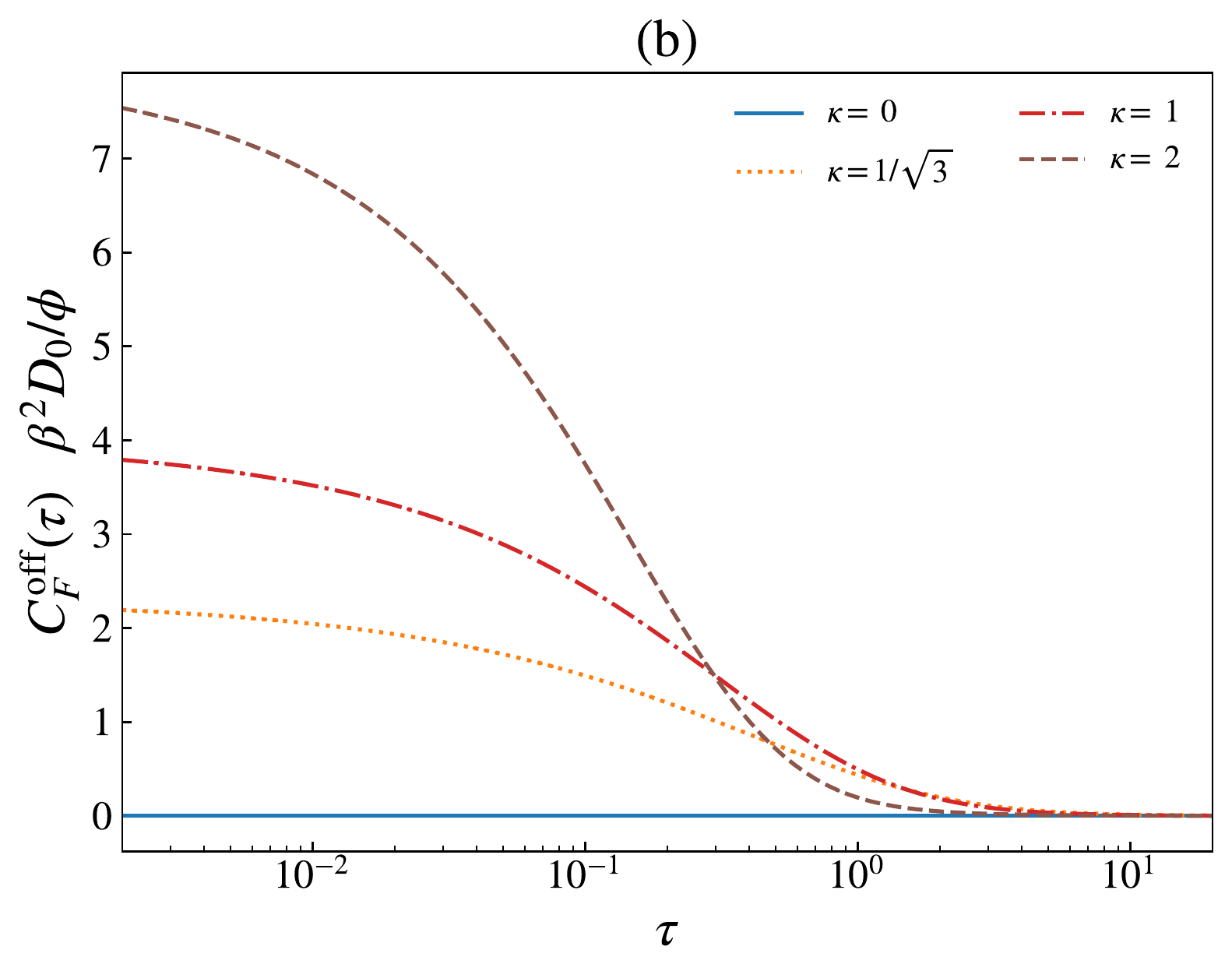}
\end{subfigure}
\caption{Diagonal and off-diagonal elements of the force autocorrelation tensor (FACT) of interacting hard disks as a function of reduced time $\tau = t/\tau_0$, where $\tau_0 = \sigma^2/(2D_0)$. (a) The diagonal elements of the FACT $C_F^\mathrm{diag}(\tau)$ (Eq.~\eqref{FF_tensor_diagonal}) are negative for a range of $\tau$ (see also inset for a finer resolution) indicating an anti-correlation of the force. For $\tau \rightarrow 0$ $C_F^\mathrm{diag}$ diverges to positive infinity, as expected for hard disks \cite{hanna1981velocity}. (b) The off-diagonal elements of the FACT $C_F^\mathrm{off}(\tau)$ (Eq.~\eqref{FF_tensor_offdiagonal}) are nonnegative and remain finite as $t\rightarrow 0$. Specifically $C_F^\mathrm{off}(\tau)$ scales linearly with $\kappa$ as $t\rightarrow 0$, which could be useful for estimating the odd-diffusion parameter in experiments~\cite{yasuda2022time}. 
}
\label{fig:FACT}
\end{figure*}

The integrals in Eq.~\eqref{R_1_integral_expresion} and Eq.~\eqref{R_minus1_integral_expresion} are evaluated in Appendix~\ref{integral_expressions_from_Gradshteyn_Ryzhik}.
The resulting force autocorrelation tensor is given as
\begin{equation}
\tilde{\mathsf{C}}_F(s) = \tilde{C}^\mathrm{diag}_F(s) \mathsf{1} + \tilde{C}^\mathrm{off}_F(s) \boldsymbol{\varepsilon},
\end{equation}
where

\begin{align}
\label{FF_tensor_diagonal}  
&\tilde{C}^\mathrm{diag}_F(s) = \frac{2\ \phi}{\beta^2 D_0}\ K_1(\sqrt{\tau_0 s}) \nonumber \\ &\quad \times \frac{\sqrt{\tau_0 s}K_0(\sqrt{\tau_0 s}) + K_1(\sqrt{\tau_0 s})}{\left(\sqrt{\tau_0 s}K_0(\sqrt{\tau_0 s}) + K_1(\sqrt{\tau_0 s})\right)^2 + \left(\kappa K_1(\sqrt{\tau_0 s})\right)^2}, \\
\label{FF_tensor_offdiagonal}
&\tilde{C}^\mathrm{off}_F(s) = \frac{2\ \kappa\phi}{\beta^2 D_0}\ K_1(\sqrt{\tau_0 s})\nonumber \\ &\quad \times \frac{K_1(\sqrt{\tau_0 s})}{\left(\sqrt{\tau_0 s}K_0(\sqrt{\tau_0 s}) + K_1(\sqrt{\tau_0 s})\right)^2 + \left(\kappa K_1(\sqrt{\tau_0 s})\right)^2},
\end{align}

are the diagonal and antisymmetric off-diagonal parts of the FACT, respectively. Note that the off-diagonal elements $\tilde{C}^\mathrm{off}_F(s)$ are proportional to the odd-diffusion parameter $\kappa$ and therefore vanish in case of normal diffusion ($\kappa = 0$). The diagonal and off-diagonal elements of the FACT are plotted in Fig.~\ref{fig:FACT} as a function of reduced time $\tau = t/\tau_0$.

It is interesting to investigate the short- and long-time asymptotics of the elements of the FACT. Therefore we define the modified Laplace variable $z = s \tau_0$ to simplify the notation. The short-time behavior of the FACF, i.e. the behavior on time-scales $t\ll\tau_0$, can analytically be obtained from Eq.~\eqref{FF_tensor_diagonal} in the limit $z\tau_0 \gg 1$, and behaves asymptotically as \cite{arfken2012mathematical}
\begin{equation}
\label{app_c_diag_short_time}
\tilde{C}^\mathrm{diag}_F(z) \sim \frac{2\phi}{D_0 \beta^2} \left[ \frac{1}{\sqrt{z}} - \frac{1}{2 z} + \mathcal{O}(z^{-5/4})\right],
\end{equation}
as $z \to \infty$. The leading order behavior at short times of $C^\mathrm{diag}_F(\tau)$ therefore is $\tau^{-1/2}$ \cite{oberhettinger1973tables}. 

The long-time behavior, i.e. $t\gg \tau_0$, can analytically be obtained from Eq.~\eqref{FF_tensor_diagonal} in the limit $z\tau_0 \ll 1$, and behaves asymptotically as \cite{arfken2012mathematical} 
\begin{subequations}
\label{c_diag_long_time}
\begin{align}
\label{app_c_diag_long_time}
\tilde{C}^\mathrm{diag}_F(z) &\sim \frac{2 \phi}{D_0\beta^2} \frac{1}{1 + \kappa^2} \Big[ 1 + A^\mathrm{diag}(\kappa)\ z \ln(z) \nonumber\\ 
&\quad + B^\mathrm{diag}(\kappa)\ z^2 \ln^2(z) + C^\mathrm{diag}(\kappa)\ z^2 \ln(z) \nonumber \\
&\quad + D^\mathrm{diag}(\kappa)\ z +\mathcal{O}(z^2)\Big] 
\end{align}
for $z \to 0$. The subsequent terms are proportional to $z^n \ln^m(z)$ with integer $n \geq 3, m \geq 0$ and therefore of infinitesimal higher order than $z^2$. The coefficients read
\begin{align}
    A^\mathrm{diag}(\kappa) &= \frac{1 -\kappa^2}{2(1 + \kappa^2)},\\
    B^\mathrm{diag}(\kappa) &= \frac{1 - 6\kappa^2 + \kappa^4}{8(1 + \kappa^2)^2},\\
    C^\mathrm{diag}(\kappa) &= \frac{1-\kappa^4 + 2(\gamma - \ln(2))(1 - 6\kappa^2 + \kappa^4)}{4(1 + \kappa^2)^2}, \\
    D^\mathrm{diag}(\kappa) &= \frac{(1 -\kappa^2)(\gamma - \ln(2))}{1 + \kappa^2}.
\end{align}
\end{subequations}
Here $\gamma = 0.5772$ is the Euler-Mascheroni constant. 
From Eq.~\eqref{app_c_diag_long_time} it can be seen that the long-time behavior of $C_F^\mathrm{diag}(\tau)$ strongly depends on $\kappa$. The leading order term is $\mathcal{O}(z \ln(z))$ and results in a decay of $C_F^\mathrm{diag}$ as $\tau^{-2}$ \cite{hull1955asymptotic} for all $\kappa$, except for $\kappa=1$. There, the prefactor vanishes, $A^\mathrm{diag}(1) =0$, and the subsequent term of $\mathcal{O}(z^2\ln^2(z))$ gives rise to a decay of $C_F^\mathrm{diag}$ as $\tau^{-3}\ln(\tau)$ \cite{hull1955asymptotic, abramowitz1968handbook}. 



The short-time behavior of $C_F^\mathrm{off}(\tau)$ can analytically be obtained from Eq.~\eqref{FF_tensor_offdiagonal} in the limit $z\tau_0\gg 1$ and behaves asymptotically as \cite{arfken2012mathematical} 
\begin{equation}
\tilde{C}^\mathrm{off}_F(z) \sim\frac{2\phi}{D_0 \beta^2}\  \kappa \left[ \frac{1}{z} - \frac{1}{z^{3/2}} + \mathcal{O}(z^{-7/4})\right]
\end{equation}
as $z \to \infty$. Therefore the leading order short-time behavior of $C^\mathrm{off}_F$ is independent of time \cite{oberhettinger1973tables} but depends linearly on $\kappa$. Such a scaling of the off-diagonal elements with $\kappa$ at short times has been recently derived by Yasuda et. al in Ref.~\cite{yasuda2022time} for odd Langevin systems. The authors also pointed out that this could be useful for estimating the odd-diffusion parameter in experiments.

The long-time behavior of $C_F^\mathrm{off}(\tau)$ can be analytically obtained from Eq.~\eqref{FF_tensor_offdiagonal} in the limit $z\tau_0 \ll 1$ and behaves asymptotically as \cite{arfken2012mathematical}
\begin{subequations}
\begin{align}
\tilde{C}^\mathrm{off}_F(z) &\sim  \frac{2\phi}{D_0 \beta^2} \frac{\kappa}{1 + \kappa^2} \big[1 + A^\mathrm{off}(\kappa)\ z\ln(z) \nonumber \\
& \quad + D^\mathrm{off}(\kappa)\ z + \mathcal{O}(z^2)\big],
\end{align}
as $z \to 0$, where the coefficients are given by 
\begin{align}
    A^\mathrm{off}(\kappa) &= \frac{1}{1+\kappa^2}, \\
    D^\mathrm{off}(\kappa) &= \frac{2(\gamma - \ln(2))}{1+\kappa^2}.
\end{align}
\end{subequations}
The leading order term is $\mathcal{O}(z \ln(z))$ and results in a decay of $C_F^\mathrm{off}$ as $\tau^{-2}$ \cite{hull1955asymptotic}, for all $\kappa$. Note that $\tilde{C}^\mathrm{off}_F$ has no leading order contributions of $\mathcal{O}(z^2 \ln^2(z))$ or $\mathcal{O}(z^2 \ln(z))$, i.e. by analogy to \eqref{c_diag_long_time}, $B^\mathrm{off} = C^\mathrm{off} = 0$.

\section{Integral expressions in the force autocorrelation}
\label{integral_expressions_from_Gradshteyn_Ryzhik}
To evaluate the integrals in $\tilde{R}_1$ and $\tilde{R}_{-1}$ in Eq.~\eqref{R_1_integral_expresion} and Eq.~\eqref{R_minus1_integral_expresion} one can use Gradshteyn and Ryzhiks \textit{Table of Integrals, Series and Products} \cite{gradshteyn2014table} and Abramowitz and Steguns \textit{Handbook of Mathematical Functions} \cite{abramowitz1968handbook}.
All integrals in $\tilde{R}_1$ and $\tilde{R}_{-1}$ can be reduced to 
\begin{align}
\mathrm{(i)} &= \int_0^\infty\mathrm{d}x\ \frac{x\ J_1^2(x)}{x^2 + b^2},
\end{align}
or
\begin{align}
\mathrm{(ii)} &= \int_0^\infty\mathrm{d}x\ \frac{x^2\ J_1(xa) J_0(xa)}{x^2 + b^2},
\end{align}
where $J_n$ are the Bessel functions of order $n$ and $a$ and $b$ are constants.
Integral (i) is (see \cite{gradshteyn2014table}, 6.535) 
\begin{equation}
\int_0^\infty\mathrm{d}x\ \frac{x\ J_1^2(x)}{x^2 + b^2} = I_1(b)\ K_1(b),
\end{equation}
where $I_n(b)$ and $K_n(b)$ are the modified Bessel functions of the first and second kind, respectively. 

Integral (ii) is not listed in \cite{gradshteyn2014table} but can be calculated straightforwardly from the formula for the $k$th derivative of a Bessel function of order $n$ (see \cite{abramowitz1968handbook}, 9.1.30), 
\begin{equation}
\left(\frac{1}{x}\ \frac{\mathrm{d}}{\mathrm{d}x}\right)^k \left( x^n\ J_n(x)\right) = x^{n-k}\ J_{n-k}(x)
\end{equation}
and the relation $J_{-n}(x) = (-1)^n\ J_n(x)$ (see \cite{abramowitz1968handbook}, 9.1.5), one obtains
\begin{equation}
\int\mathrm{d}x\ J_0(x)\ J_1(x) = - \frac{J_0(x)^2}{2} + const.
\end{equation}
This together with a relation for the Wronskian $\mathcal{W}[K_n(x), I_n(x)]$ (see \cite{abramowitz1968handbook}, 9.6.15)
\begin{equation}
\label{Wronskian_expression}
\mathcal{W}[K_n(x), I_n(x)] = I_n(x)\ K_{n+1}(x) + I_{n + 1}(x)\ K_n(x) = \frac{1}{x}
\end{equation}
can be used to evaluate integral (ii) by partial integration.
This gives
\begin{equation}
\int_0^\infty\mathrm{d}x\ \frac{x^2\ J_1(ax) J_0(ax)}{x^2 + b^2} = b\ I_0(ab)\ K_1(ab) - \frac{1}{2a}.
\end{equation}

Using the integrals (i) and (ii) in the expressions for $\tilde{R}_1$ and $\tilde{R}_{-1}$ in Eq.~\eqref{R_1_integral_expresion} and Eq.~\eqref{R_minus1_integral_expresion}, together with
another use of the Wronskian expression Eq.~\eqref{Wronskian_expression},
one arrives at the form of the diagonal and off-diagonal parts of the FACT, $\tilde{C}^\mathrm{diag}_F(s)$ and $\tilde{C}^\mathrm{off}_F(s)$, in
Eq.~\eqref{FF_tensor_diagonal} and Eq.~\eqref{FF_tensor_offdiagonal}, respectively.

\bibliographystyle{unsrt}